\documentclass[10pt]{amsart}
\usepackage{amsmath,amsfonts,amscd,amssymb,epsf}
\newtheorem{theorem}{Theorem}[section]
\newtheorem{proposition}[theorem]{Proposition}
\newtheorem{lemma}[theorem]{Lemma}
\newtheorem{corollary}[theorem]{Corollary}
\theoremstyle{definition}

\theoremstyle{remark} 
\numberwithin{equation}{section}

\DeclareMathOperator{\id}{id} 
 
 \DeclareMathOperator{\Homeo}{Homeo}

\DeclareMathOperator{\PSL}{PSL} 

\newcommand{\Z}{{\mathbb{Z}}}\newcommand{\Del}{\mathbb{D}}

\newcommand{\C}{{\mathbb{C}}}
\newcommand{\R}{{\mathbb{R}}}
\newcommand{\mL}{\mathcal{L}}
\newcommand{\B}{\mathcal{B}}
\newcommand{\mM}{\mathcal{M}}
\newcommand{\F}{\mathcal{F}}
\newcommand{\mC}{\mathcal{C}}
\newcommand{\bk}{\backslash}
\newcommand{\pa}{\partial}

\newcommand{\ov}{\overline}

\newcommand{\vep}{\varepsilon}
\newcommand{\z}{\bar{z}}

\begin{document}
\title{Conformal Mappings and Dispersionless Toda hierarchy}
\author{ Lee-Peng Teo}\address{Faculty of Information Technology \\
Multimedia University \\ Jalan Multimedia, Cyberjaya\\
63100, Selangor\\Malaysia}  \email{lpteo@mmu.edu.my}

\begin{abstract}Let $\mathfrak{D}$ be the space consists of pairs   $(f,g)$, where $f$ is a univalent function on the unit disc with $f(0)=0$,   $g$ is a univalent function on the exterior of the unit disc with $g(\infty)=\infty$ and $f'(0)g'(\infty)=1$. In this article, we define the time variables $t_n, n\in \Z$, on $\mathfrak{D}$ which are holomorphic with respect to the natural complex structure on $\mathfrak{D}$ and can serve as local complex coordinates for $\mathfrak{D}$. We show that the evolutions of the pair $(f,g)$ with respect to these time coordinates are governed by the dispersionless Toda hierarchy flows.  An explicit tau function is constructed for the dispersionless Toda hierarchy. By restricting $\mathfrak{D}$ to the subspace $\Sigma$ consists of pairs where $f(w)=1/\overline{g(1/\bar{w})}$, we obtain the integrable hierarchy of conformal mappings considered by Wiegmann and Zabrodin \cite{WZ}. Since every $C^1$ homeomorphism $\gamma$ of the unit circle corresponds uniquely to an element $(f,g)$ of $\mathfrak{D}$ under the conformal welding $\gamma=g^{-1}\circ f$, the space $\text{Homeo}_{C}(S^1)$ can be naturally  identified as a subspace of $\mathfrak{D}$ characterized by $f(S^1)=g(S^1)$. We show that we can naturally define complexified vector fields $\pa_n, n\in \Z$ on $\text{Homeo}_{C}(S^1)$ so that the evolutions of $(f,g)$ on $\text{Homeo}_{C}(S^1)$ with respect to $\pa_n$ satisfy the dispersionless Toda hierarchy. Finally, we show that there is a similar integrable structure for the Riemann mappings $(f^{-1}, g^{-1})$. Moreover, in the latter case, the time variables are Fourier coefficients of $\gamma$ and $1/\gamma^{-1}$.
\end{abstract}
 \maketitle

\section{Introduction}

Introduced in \cite{TT5, TT1} as the dispersionless limit of the well-known Toda lattice hierarchy \cite{UT}, dispersionless Toda hierarchy can also be
interpreted as describing the evolutions of the coefficients of two
formal power series $(\mL, \tilde{\mL})$ with respect to a set of
formal time variables $t_n$, $n\in \Z$. Here $\mL(w)= w+$ (lower
power terms), and $\tilde{\mL}(w)= w$+ higher power terms. Under certain analytic conditions, $\mL$ is a function univalent in
a neighborhood of $\infty$, and $\tilde{\mL}$ is a function
univalent in a neighborhood of the origin. It is therefore natural to link up evolutions of conformal mappings with dispersionless Toda hierarchy.
Starting from the work of Wiegmann and Zabrodin \cite{WZ}, the
integrable structure of conformal mappings has aroused
considerable interest \cite{KKMWZ, LT, MWZ, 1, 2, 3, 4, 5, 6, 7, 8, 9, 10, 11, 12, 13}. Wiegmann and Zabrodin \cite{WZ} defined a set of time variables $t_n, n\geq 0$, on the space of conformal mappings $g$  that map the exterior of the unit disc onto the exterior of a simply connected domain that contains the origin. They showed that the evolutions of the conformal mappings $(g(w), 1/\ov{g(1/\bar{w})})$ with respect to $(\ldots, -\bar{t}_2, -\bar{t}_1, t_0, t_1, t_2, \ldots)$ satisfy the dispersionless Toda hierarchy.  They also
defined the notion of tau function for analytic curves, which is
the tau function for the hierarchy. Later it was
revealed that this problem  is   closely related to the Dirichlet boundary
problem and two dimensional inverse potential problem \cite{MWZ,
Z}, and it can be put under the framework of conformal field theory \cite{LT}.    By a straightforward modification, it
was shown in \cite{MWZ} that the deformation of the  conformal
mapping $f$ of the interior domain  can also be described by the dispersionless Toda
hierarchy. However,   the evolutions of the interior
mappings $f$ and the evolutions of the exterior mappings $g$ are treated  using
different time coordinates.

In this paper, we consider a more general conformal mappings problem.  Denote by $\mathfrak{D}$ the space consists of pairs of conformal mappings $(f,g)$, where $f$ is a univalent function on the unit disc $\Del$ and $g$ is a univalent function on the exterior of the unit disc $\Del^*$, normalized so that $f(0)=0$, $g(\infty)=\infty$ and $f'(0)g'(\infty)=1$. We also assume that   both $f$ and $g$ can be extended to $C^1$ homeomorphisms of the extended complex plane. Moreover, the interior domain $\Omega_1^+=f(\Del)$ does not contain $\infty$ and the exterior domain $\Omega_2^-=g(\Del^*)$ does not contain the origin. A set of complex time variables $t_n, n\in \Z$, are defined on $\mathfrak{D}$ so that the coefficients of $f$ and $g$ depend holomorphically on $t_n$. We construct a tau function on the space $\mathfrak{D}$ and use it to show that the evolutions of the conformal mappings $(g,f)$ with respect to the set of time variables $t_n, n\in \Z$, satisfy the dispersionless Toda hierarchy, with $\mL=g$ and $\tilde{\mL}=f$.

 In the language of Takasaki and Takebe \cite{TT5}, the solution to the dispersionless Toda hierarchy we considered is the solution to the Riemann-Hilbert problem
\begin{equation}\label{eq4_30_1}
\mL \mM^{-1}=\tilde{\mL}, \hspace{1cm}\mM=\tilde{\mM},
\end{equation}which has been considered in the formal level in \cite{TT3} in relation to two dimensional string theory. Here $\mM$ and $\tilde{\mM}$ are the Orlov-Shulman functions. A consequence of \eqref{eq4_30_1} is that $\mL$ and $\tilde{\mL}$ satisfy the string equation $\left\{\mL, \tilde{\mL}^{-1}\right\}_T=1$, or more precisely,
$$\left\{g(w), f(w)^{-1}\right\}_T = w\frac{\pa g(w)}{\pa w}\frac{\pa f(w)^{-1}}{\pa t_0} -w\frac{\pa f(w)^{-1}}{\pa w}\frac{\pa g(w)}{\pa t_0}=1.$$

Let $\Sigma$ be the subspace of $\mathfrak{D}$ consists of $(f,g)$ with $f(w)=1/\overline{g(1/\bar{w})}$. We show that $\Sigma$ is characterized by $\bar{t}_n =-t_{-n}$. The restriction of the dispersionless Toda flows to the subspace $\Sigma$ is   the integrable structure of conformal mappings considered by Wiegmann and Zabrodin \cite{WZ}. Another interesting subspace of $\mathfrak{D}$ is the space characterized by $f(S^1)=g(S^1)$, which is equivalent to $g^{-1}\circ f$ being a $C^1$ homeomorphism of the unit circle. In fact, for every $C^1$ homeomorphism $\gamma$ of the unit circle, there is a unique element $(f,g)$ of $\mathfrak{D}$ such that $\gamma=g^{-1}\circ f$. Therefore, we can identify the space $\text{Homeo}_{C}(S^1)$ of $C^1$ homeomorphisms of the unit circle as a subspace of $\mathfrak{D}$ containing $(f,g)$ with $f(S^1)=g(S^1)$. Assume that  this subspace is locally defined by the equations $\bar{t}_n = \mathcal{Z}_n (t_m)$, then we can define vector fields $\pa/\pa \mathrm{t}_n, n\in \Z$, as a restriction of $\pa/\pa t_n, n\in \Z$, to $\text{Homeo}_{C}(S^1)$. It is easy to deduce from the results on $\mathfrak{D}$ that the evolutions of $(g,f)\in \text{Homeo}_C(S^1)$ with respect to $\mathrm{t}_n, n\in \Z$, are also governed by the dispersionless Toda flows.

The inverse functions $f^{-1}, g^{-1}$ are Riemann mappings of the respective domains $\Omega_1^+$ and $\Omega_2^-$. It is also interesting to study their evolutions under the context of integrable hierarchies. We show that restricted to the space $\text{Homeo}_{C}(S^1)$, we can define   time variables $\mathsf{t}_n, n\in \Z$, which are some Fourier coefficients of $\gamma$ and $1/\gamma^{-1}$. There are complexified vector fields $\pa_n$ on $\text{Homeo}_C(S^1)$ whose action on $\mathsf{t}_n, n\in \Z$, is given by $\pa_n t_m=\delta_{n,m}$. Therefore, we can identify   $\pa/\pa\mathsf{t}_n$ on $\text{Homeo}_C(S^1)$ with $\pa_n$. We construct a tau function on $\text{Homeo}_C(S^1)$ and show that the evolutions of $(g^{-1}, f^{-1})$ with respect to $\mathsf{t}_n, n\in\Z,$ satisfy the dispersionless Toda hierarchy.

The layout of this papers is as follows. In Section 2, we review some facts we need about generalized Grunsky coefficients, generalized Faber polynomials and dispersionless Toda hierarchy. In Section 3, we prove that there is an   integrable structure on the space $\mathfrak{D}$ of pairs of conformal mappings. In Section 4, we discuss the Riemann-Hilbert data associated to our solution to the dispersionless Toda hierarchy. In Section 5, we discuss the relation of our work with the work of Wiegmann and Zabrodin. In Section 6, we consider the restriction of the integrable hierarchy to conformal mappings $(f,g)$ satisfying $f(S^1)=g(S^1)$. In Section 7, we consider the corresponding problem for Riemann mappings.

\section{Background Materials}
\subsection{Grunsky coefficients and Faber polynomials}
We review some concepts we need about  univalent functions. For
details, see \cite{Pom, Duren, Teo03, Teo04}.

Let $\mathfrak{F}(z) = \alpha_1z +\alpha_2 z^2 + \ldots$ be a function
univalent in a neighborhood of the origin and $\mathfrak{G}(z) =
\beta z + \beta_0 + \beta_1 z^{-1} + \ldots$, $\beta=\alpha_1^{-1}$, be a function
univalent in a neighborhood of $\infty$. We define the
generalized Grunsky coefficients $b_{m,n}$, $m,n\in\Z$ and Faber
polynomials $P_n$ and $Q_n$ by the following formal power series
expansion:
\begin{equation*}\begin{split}
\log \frac{\mathfrak{G}(z)-\mathfrak{G}(\zeta)}{z-\zeta}&=
\log \beta-\sum_{m=1}^{\infty}\sum_{n=1}^{\infty} b_{mn} z^{-m}\zeta^{-n},\\
\log \frac{\mathfrak{G}(z) - \mathfrak{F}(\zeta)}{z} &= \log
\beta-\sum_{m=1}^{\infty}
\sum_{n=0}^{\infty} b_{m,-n} z^{-m} \zeta^n,\\
\log \frac{\mathfrak{F}(z)-\mathfrak{F}(\zeta)}{z-\zeta} &=
-\sum_{m=0}^{\infty}
\sum_{n=0}^{\infty} b_{-m, -n} z^{m} \zeta^n,\\
\log \frac{\mathfrak{G}(z) -w}{bz} &= -\sum_{n=1}^{\infty}
\frac{P_n(w)}{n}
z^{-n} ,\\
\log \frac{w- \mathfrak{F}(z)}{w} &= \log
\frac{\mathfrak{F}(z)}{\alpha_1 z}-\sum_{n=1}^{\infty} \frac{Q_n(w)}{n}
z^{n};\end{split}
\end{equation*}
and for $m\geq 0$, $n\geq 1$, $b_{-m,n}=b_{n,m}$. By definition,
the Grunsky coefficients are symmetric, i.e., $b_{m,n} =b_{n,m}$
for all $m,n\in\Z$. The coefficient $b_{0,0}$ is given explicitly by $-\log \alpha_1=\log \beta$, where $\alpha_1= \mathfrak{F}'(0)$ and $\beta=g'(\infty)$. $P_n(w)$ is a polynomial of degree $n$ in $w$
and $Q_n(w)$ is a polynomial of degree $n$ in $1/w$. More
precisely,
\[
P_n(w) = (\mathfrak{G}^{-1}(w)^n)_{\geq 0}, \hspace{1cm}Q_n(w) =
\left(\frac{1}{\mathfrak{F}^{-1}(w)^{n}}\right)_{\leq 0}.\] Here when $S$ is a subset
of integers and $A(w)=\sum_{n} A_nw^n$ is a (formal) power series,
we denote by $(A(w))_{S}$ the truncated sum $\sum_{n\in S} A_n w^n$.

The functions $\log (\mathfrak{G}(z)/z)$, $P \circ \mathfrak{G}$
and $Q\circ \mathfrak{G}$ are meromorphic in a neighborhood of
$\infty$ and the functions $\log (\mathfrak{F}(z)/z)$, $P_n\circ
\mathfrak{F}$ and $Q_n \circ \mathfrak{F}$ are meromorphic in a
neighborhood of the origin. Their power series expansions are
given by
\begin{equation}\label{iden1}\begin{split}
\log\frac{\mathfrak{G}(z)}{z} &= \log \beta-\sum_{m=1}^{\infty}
b_{0,m} z^{-m},\hspace{1.5cm} \log\frac{\mathfrak{F}(z)}{z} =\log
\alpha_1-\sum_{m=1}^{\infty}
b_{0,-m} z^{m}\\
P_n (\mathfrak{G}(z)) &= z^n + n\sum_{m=1}^{\infty} b_{nm}
z^{-m},\hspace{1.5cm}P_n (\mathfrak{F}(z)) =
nb_{n,0}+n\sum_{m=1}^{\infty}
b_{n, -m} z^m,  \\
Q_n(\mathfrak{G}(z)) &= -nb_{-n,0} + n\sum_{m=1}^{\infty}
b_{-n,m} z^{-m},\hspace{0.4cm}Q_n (\mathfrak{F}(z))=
z^{-n} + n\sum_{m=1}^{\infty} b_{-n,-m} z^{m}.\end{split}
\end{equation}

\subsection{Dispersionless Toda hierarchy}
The dispersionless Toda hierarchy is a hierarchy of equations
describing the evolutions of the coefficients of a pair of formal
power series $(\mL, \tilde{\mL})$, where
\begin{equation}\label{series}\begin{split}
\mL(w) = r(\boldsymbol{t})w + \sum_{n=0}^{\infty} u_{n+1}(\boldsymbol{t}) w^{-n},\\
(\tilde{\mL}(w))^{-1} = r(\boldsymbol{t})w^{-1} +
\sum_{n=0}^{\infty} \tilde{u}_{n+1}(\boldsymbol{t})
w^{n}\nonumber.
\end{split}\end{equation}
Here $r(\boldsymbol{t})$, $u_n(\boldsymbol{t})$ are functions of
$t_n$, $n\in \Z$, which we denote collectively by
$\boldsymbol{t}$; $w$ is a formal variable independent of
$\boldsymbol{t}$. The evolution of the coefficients $u_n$ are
encoded in the following Lax equations:
\begin{equation}\label{Lax}\begin{split}
\frac{\pa \mL}{\pa t_n} =\{ \B_n, \mL\}_T, \hspace{2cm}
\frac{\pa \mL}{\pa t_{-n}} = \{\tilde{B}_n, \mL\}_T,\\
\frac{\pa \tilde{\mL}}{\pa t_n} =\{ \B_n, \tilde{\mL}\}_T,
\hspace{2cm} \frac{\pa \tilde{\mL}}{\pa t_{-n}} = \{\tilde{B}_n,
\tilde{\mL}\}_T.\end{split}
\end{equation}
Here $\{\cdot, \cdot\}_T$ is the Poisson bracket
\begin{equation*}
\{ f, g\}_T = w \frac{\pa f}{\pa w} \frac{\pa g}{\pa
t_0}-w\frac{\pa f}{\pa t_0} \frac{\pa g}{\pa w},
\end{equation*}
and \begin{equation*}
\B_n=(\mL^n)_{>0}+\frac{1}{2}(\mL^n)_0,\hspace{1cm}\tilde{B}_n=(\tilde{\mL}^n)_{<0}
+\frac{1}{2}(\tilde{\mL}^n)_{0}.\end{equation*} Proposition 3.1 in
\cite{Teo03} can be reformulated as
\begin{proposition}\label{Hirota} If there exists a function $\tau$ of
$\boldsymbol{t}$ such that $ \log\tau$   generates the generalized Grunsky
coefficients of a pair $(\mathfrak{F}, \mathfrak{G})$ of formal
power series, namely
\begin{equation*}
\frac{\pa^2\\log\tau(\boldsymbol{t})}{\pa t_m\pa t_n}
=\begin{cases}-|mn|b_{m,n}(\boldsymbol{t}),\hspace{1cm}&\text{if}\;\;m\neq 0, n\neq 0\\
|m| b_{m,0}(\boldsymbol{t}), &\text{if}\;\;m\neq 0,n= 0\\
-2b_{0,0}(\boldsymbol{t}),&\text{if}\;\;m=n=0,
\end{cases}
\end{equation*}then the pair of
formal power series $( \mathfrak{G}^{-1}, \mathfrak{F}^{-1})$
satisfies the dispersionless Toda hierarchy. Here $
\mathfrak{G}^{-1}$ and  $\mathfrak{F}^{-1}$ are the inverse
functions of $ \mathfrak{G}$ and $\mathfrak{F}$
respectively.
\end{proposition}

\section{Pairs of conformal mappings and  dispersionless Toda hierarchy}\label{ext}

Let $\mathbb{D}$ be the unit disc and $\Del^*$ its exterior. In this section, we consider the space of pairs of conformal mappings and show that it has an integrable structure modeled by dispersionless Toda hierarchy. First we define the spaces:
\begin{equation*}\begin{split}
\mathfrak{S}_0=\Bigl\{ &f: \mathbb{D} \rightarrow \C \;\text{univalent}
\;\bigr\vert\;
f(w)=a_1 w+a_2w^2+ \ldots; a_1\neq0; \\
& \infty \notin f(\Del);\;\;f \;\text{is extendable to a $C^1$   homeomorphism
of $\hat{\C}$}.\Bigr\},\\
\mathfrak{S}_{\infty}=\Bigl\{ &g: \mathbb{D}^* \rightarrow \C \;\text{univalent}
\;\bigr\vert\;
g(w)=b w+b_0 + b_1w^{-1}+\ldots; b\neq0; \\
&0\notin g(\Del^*);\;\;g \;\text{is extendable to a $C^1$   homeomorphism
of $\hat{\C}$}.\Bigr\},\\
\mathfrak{D}=&\left\{(f,g)\; \bigr\vert\; f\in \mathfrak{S}_0, g\in \mathfrak{S}_{\infty};\;
f'(0)g'(\infty) = a_1b=1 .\right\}.\end{split}
\end{equation*}Let $\Omega_1^+= f(\mathbb{D})$ and $\Omega_1^-$ its exterior,
$\Omega_2^-=g(\mathbb{D}^*)$ and $\Omega_2^+$ its exterior. $\mathcal{C}_1$ and $\mathcal{C}_2$ denotes the $C^1$ curves $\mathcal{C}_1=f(S^1)$ and $\mathcal{C}_2=g(S^1)$ respectively. Notice that $f^{-1}$ is a Riemann mapping of $\Omega^+_1$ and $g^{-1}$ is a Riemann mapping of $\Omega_2^-$.
The functions $t_n$ and $v_n$ are functions on the space $\mathfrak{D}$  defined in the following way: For $n\geq 1$,
\begin{equation}\label{eq4_12_6}\begin{split}
t_n=&\frac{1}{2\pi i
n}\oint_{S^1}\frac{g(w)^{-n}}{f(w)}dg(w)=\frac{1}{2\pi i
n}\oint_{\mathcal{C}_2}\frac{z^{-n}}{f\circ g^{-1}(z)}dz,\\t_{-n}=&
\frac{-1}{2\pi in}\oint_{S^1} g(w) f(w)^{n-2} df(w)=\frac{-1}{2\pi in}\oint_{\mathcal{C}_1}z^{n-2}g\circ f^{-1}(z) dz,
\\
v_n =&\frac{1}{2\pi i }\oint_{S^1}\frac{g(w)^n}{f(w)}dg(w)=\frac{1}{2\pi i
}\oint_{\mathcal{C}_2}\frac{z^{n}}{f\circ g^{-1}(z)}dz,
\\v_{-n} =&\frac{-1}{2\pi i}\oint_{S^1} g(w)
f(w)^{-n-2} df(w)=\frac{-1}{2\pi i}\oint_{\mathcal{C}_1}z^{-n-2}g\circ f^{-1}(z) dz. \end{split}\end{equation}For $n=0$,\begin{equation}\label{eq4_12_7}\begin{split}
t_0 =&\frac{1}{2\pi i
}\oint_{S^1}\frac{1}{f(w)}dg(w)=\frac{1}{2\pi i} \oint_{S^1}\frac{g(w)}{f(w)^2} df(w), \\
=&\frac{1}{2\pi i
}\oint_{\mathcal{C}_2}\frac{1}{f\circ g^{-1}(z)}dz=\frac{1}{2\pi i}\oint_{\mathcal{C}_1}z^{-2}g\circ f^{-1}(z) dz,\\
v_0=&\frac{1}{2\pi i} \oint_{S^1}\left\{ \log\frac{g(w)}{w}
\frac{ g'(w)}{f(w)}-  \log
\frac{f(w)}{w} g(w)
\frac{f'(w)}{f(w)^2}- \frac{1}{w}\frac{g(w)}{f(w)}\right\}
dw.
\end{split}\end{equation}
It is obvious that $t_n$ and $v_n$ depend analytically on the coefficients of the functions $f$ and $g$.  We also
 define the following functions:
\begin{equation}\label{eq12_29_15}\begin{split}
S_{\pm}(z) =\frac{1}{2\pi i} \oint_{S^1} \frac{(1/f)(w)
g'(w)}{g(w)-z} dw=\frac{1}{2\pi i} \oint_{\mathcal{C}_2}\frac{1/(f\circ g^{-1})(\zeta)}{\zeta-z}d\zeta, \hspace{1cm} z\in \Omega_2^{\pm},\\
\tilde{S}_{\pm}(z)=\frac{1}{2\pi i} \oint_{S^1} \frac{g(w)
f'(w)}{f(w)^2(f(w)-z)} dw=\frac{1}{2\pi i} \oint_{\mathcal{C}_1}\frac{g\circ f^{-1}(\zeta)}{\zeta^2(\zeta-z)}d\zeta,\hspace{1cm} z\in \Omega_1^{\pm} .\end{split}
\end{equation}
They are meromorphic functions in the respective domains. In a
neighborhood of the origin or $\infty$, they have the series
expansion
\begin{equation}\label{eq4_12_1}\begin{split}
S_+(z) &= \sum_{n=1} nt_n z^{n-1}, \hspace{2cm} \tilde{S}_{+} (z)
= -\sum_{n=1}^{\infty} v_{-n} z^{n-1},\\
S_-(z) &=-t_0 z^{-1} -\sum_{n=1}^{\infty}
v_nz^{-n-1},\hspace{0.5cm} \tilde{S}_{-}(z) = -t_0z^{-1}
+\sum_{n=1}^{\infty} nt_{-n} z^{-n-1}.
\end{split}\end{equation}
Moreover, restricted to the curve $\mathcal{C}_2$,
\begin{equation}\label{eq4_28_1}
\frac{1}{f\circ g^{-1}(z)}=S_+(z)-S_-(z)=\sum_{n=1} nt_n z^{n-1}+t_0 z^{-1} +\sum_{n=1}^{\infty}
v_nz^{-n-1};
\end{equation}whereas restricted to the curve $\mathcal{C}_1$,
\begin{equation}\label{eq4_28_2}
\frac{g\circ f^{-1}(z)}{z^2} = \tilde{S}_+(z)-\tilde{S}_-(z)=-\sum_{n=1}^{\infty} nt_{-n} z^{-n-1}+t_0 z^{-1}-\sum_{n=1}^{\infty} v_{-n} z^{n-1}.
\end{equation}

We would like to show that $\{t_n ,
 n\in \Z\}$ is a complete set of local   complex coordinates on the space $\mathfrak{D}$ and therefore the partial derivatives $\frac{\pa}{\pa t_n}, \frac{\pa }{\pa \bar{t}_n}$ are well-defined. First we have the following lemmas.
\begin{lemma}\label{lemma1}
Let $t\mapsto (f_t, g_t)$ be any smooth curve  on $\mathfrak{D}$, then
\begin{equation*}
\frac{(\pa  (g_t\circ f_t^{-1})/\pa t)\circ f_t(w)}{f_t(w)^2}f_t'(w) =
\left(\frac{\pa}{\pa t} \frac{1}{f_t\circ g_t^{-1}}\right)\circ g_t(w)
g_t'(w), \hspace{1cm} w\in S^1.
\end{equation*}
\end{lemma}
\begin{proof}
It is straightforward to verify that
\begin{equation}\label{eq4_12_5}
\begin{split}
\left(\frac{d}{dt}\frac{1}{f\circ g^{-1}}\right)(z) =& \left(\frac{-\frac{df}{dt}+\frac{f'}{g'} \frac{dg}{dt}}{f^2}\right)\circ g^{-1}(z),\\
\frac{1}{z^2}\frac{d(g\circ f^{-1})}{dt}(z) =& \left(\frac{\frac{dg}{dt}-\frac{g'}{f'} \frac{df}{dt}}{f^2}\right)\circ f^{-1}(z).
\end{split}
\end{equation}The assertion follows immediately.

\end{proof}
\begin{lemma}\label{l4_12_1}
Let $\mathcal{C}_t$, $t\in (-\vep, \vep)$ be a smooth family of
 $C^1$ curves and  let $h(z, \z, t)$ be a smooth family of $C^1$ functions defined
in a neighborhood of $\mathcal{C}_t$. Then
\begin{equation}\label{calculus}
\frac{d}{dt}\Bigr\vert_{t=0} \left(\oint_{\mathcal{C}_t} h(z,
\z,t) dz\right)=\oint_{\mathcal{C}} \frac{dh}{dt}(z,
\z,t)dz+\frac{\pa h}{\pa \z}(z,\z,t)\left(\frac{\pa \ov{w_t}}{\pa
t} dz-\frac{\pa w_t}{\pa t} d\z\right)\Bigr\vert_{t=0}
\end{equation}
Here $w_t(z,\z)$ is a family of $C^1$ functions such that
$w_0(z,\z)=\id$, $w_t(\mathcal{C}_t) = \mathcal{C}_0$. \end{lemma} The proof of this lemma is  straightforward.  In the
following, we are mostly dealing with function $h$ such that
$h_{\bar{z}}=0$ on $\mathcal{C}$. In this case the second term in \eqref{calculus} vanishes.

With these lemmas, we can show that
\begin{proposition}\label{Prop1}
If $(f_t, g_t)$, $t\in (-\vep, \vep) \subseteq\R$, is a smooth curve on
$\mathfrak{D}$ such that $dt_n/dt=0$ for all $n$, then
$(f_t, g_t) =(f_0, g_0)$ for all $t$.
\end{proposition}
\begin{proof}

If $(f_t, g_t)$ is a one parameter family in $\mathfrak{D}$ such
that $dt_n/dt=0$ for all $n$, then  \eqref{eq4_12_1}  gives
\begin{equation*}
\frac{d S_{+}}{dt}(z)=0,\hspace{2cm} \frac{d
\tilde{S}_{-}}{dt}(z)=0;
\end{equation*}
and
$$\frac{dS_-}{dt}=O(z^{-2})\hspace{1cm}\text{as}\;z\rightarrow \infty.$$ We conclude from \eqref{eq4_28_1} that
$d((1/f)\circ g^{-1})/dt$ is the boundary value of the holomorphic
function $d S_{-}/dt$ on $\Omega_2^-$, which vanish at $\infty$.
Since $g$ is holomorphic on $\mathbb{D}^*$,
\begin{equation}\label{eq4_12_2}
\frac{d}{dt}\left(\frac{1}{f\circ g^{-1}}\right)\circ g(w)
g'(w),\hspace{1cm} w\in S^1\end{equation} is the boundary value of a
holomorphic function on $\mathbb{D}^*$. Similarly, we conclude
that
\begin{equation}\label{eq4_12_3}
\frac{\left(d(g\circ f^{-1})/dt\right)\circ f(w)}{(f(w))^2}f'(w),
\hspace{1cm} w\in S^1
\end{equation}
is the boundary value of a holomorphic function on $\mathbb{D}$.
Lemma \ref{lemma1}   implies that the function holomorphic in $\Del^*$ which is equal to \eqref{eq4_12_2} on $S^1$ and the function holomorphic  in $\Del$ which is equal to \eqref{eq4_12_3} on $S^1$ agree on $S^1$. Therefore, we have a holomorphic
function on $\hat{\C}$ which vanishes at $\infty$. This function
must  be identically zero. Therefore $d(g\circ f^{-1})/dt=0$ and working
this formula out explicitly, we have
\begin{equation*}
\frac{1}{g'(w)}\frac{dg(w)}{dt}=\frac{1}{f'(w)}\frac{df(w)}{dt}, \hspace{1cm}w\in S^1.
\end{equation*}
However, restricted to $S^1$,
\begin{equation*}\begin{split}
\frac{1}{g'(w)}\frac{dg(w)}{dt}=&\frac{d\log b}{dt} w+ \text{lower
order terms in $w$.}\\
\frac{1}{f'(w)}\frac{df(w)}{dt}=&\frac{d\log a_1}{dt}
w+\text{higher order terms in $w$}.\end{split}
\end{equation*}
Comparing coefficients and using the fact that $a_1=b^{-1}$, we
conclude that $dg/dt=df/dt=0$. This implies that $(f_t, g_t)=(f_0,g_0)$ for all $t$.
\end{proof}

Proposition \ref{Prop1} shows that the set $\{t_n, n\in\Z\}$  of variables on $\mathfrak{D}$ is complete, i.e., any nontrivial vector field on $\mathfrak{D}$ will change at least one of $t_n$. To show that this set of variables are independent, we first introduce some notations. Let $\kappa_{m,n}$ and $\mathcal{P}_n, \mathcal{Q}_n$ be the generalized Grunsky coefficients and Faber polynomials of the pair of univalent functions $(f^{-1}, g^{-1})$. We have in particular $\kappa_{0,0}=-\log (f^{-1})'(0))=\log f'(0)=\log a_1=-\log b$.  The following proposition shows that the  variables $t_n, n\in\Z,$ are independent.

\begin{proposition}\label{p4_12_1}
There are variations $\pa_n, n\in \Z,$ of $(f,g)$ on $\mathfrak{D}$ so that $\pa_n t_m =\delta_{n,m}$.
\end{proposition}
\begin{proof}
For $n\geq 1$ and $w\in S^1$, let
\begin{equation*}
\begin{split}
u_n(w) =& \frac{\mathcal{P}_n'(w) f(w)^2}{f'(w)g'(w)}=\sum_{m=-\infty}^{\infty} u_{n;m}w^m,\\
u_{-n}(w) =& \frac{\mathcal{Q}_n'(w) f(w)^2}{f'(w)g'(w)}=\sum_{m=-\infty}^{\infty} u_{-n;m}w^m,
\end{split}
\end{equation*}and let
\begin{equation*}
u_0(w) =\frac{  f(w)^2}{w f'(w)g'(w)}=\sum_{m=-\infty}^{\infty} u_{0;m}w^m.
\end{equation*}Notice that since $\mathcal{P}_n'(w)$ is a  polynomial in $w$ of degree $n-1$ and $Q_n'(w)$ is a polynomial of degree $n+1$ in $w^{-1}$ without $w^{-1}$ and $w^0$ terms, the $\Z\times \Z$ matrix $\{u_{n;m}\}$ is nonsingular.  Define independent variations $\pa_n, n\in \Z$ on $\mathfrak{D}$ so that
\begin{equation*}\begin{split}
\pa_n f(w) = -f'(w)\left(\frac{1}{2}u_{n;1}w +\sum_{m=2}^{\infty} u_{n;m}w^m\right),\\
\pa_n g(w)=g'(w)\left(\frac{1}{2}u_{n;1}w +\sum_{m=0}^{\infty} u_{n;-m}w^{-m}\right).
\end{split}
\end{equation*}It follows from \eqref{eq4_12_5} that for $n\geq 1$ and $z\in \mathcal{C}_2$,
\begin{equation*}
\begin{split}
\pa_n \left(\frac{1}{f\circ g^{-1}}\right)(z) =& \left( \frac{u_n f'}{f^2}\right)\circ g^{-1}(z) = \mathcal{P}_n' \circ g^{-1} (z) (g^{-1})'(z),\\
\pa_{-n}\left(\frac{1}{f\circ g^{-1}}\right)(z) =& \left( \frac{u_{-n} f'}{f^2}\right)\circ g^{-1}(z) = \mathcal{Q}_n'  \circ g^{-1}(z)(g^{-1})'(z), \\
\pa_0 \left(\frac{1}{f\circ g^{-1}}\right)(z) =& \left( \frac{u_0 f'}{f^2}\right)\circ g^{-1}(z) = \frac{(g^{-1})'(z)}{g^{-1}(z)}.\end{split}\end{equation*}For $z\in\mathcal{C}_1$,\begin{equation*}\begin{split}
\frac{1}{z^2}\pa_n \left( g\circ f^{-1}\right)(z) =& \left( \frac{u_n g'}{f^2}\right)\circ g^{-1}(z) =\mathcal{P}_n' \circ f^{-1} (z) (f^{-1})'(z),\\
\frac{1}{z^2}\pa_{-n} \left( g\circ f^{-1}\right)(z) =& \left( \frac{u_{-n} g'}{f^2}\right)\circ g^{-1}(z) =\mathcal{Q}_n'  \circ f^{-1}(z)(f^{-1})'(z),\\\frac{1}{z^2}\pa_{0} \left( g\circ f^{-1}\right)(z) =& \left( \frac{u_{0} g'}{f^2}\right)\circ g^{-1}(z) =\frac{(f^{-1})'(z)}{f^{-1}(z)}.
\end{split}
\end{equation*}As follows from \eqref{iden1}, in a neighbourhood of $z=\infty$,
\begin{equation*}
\begin{split}
\mathcal{P}_n' \circ g^{-1}(z)(g^{-1})'(z)=&nz^{n-1}-n\sum_{m=1}^{\infty} m\kappa_{nm}z^{-m-1},\\
\mathcal{Q}_n' \circ g^{-1}(z)(g^{-1})'(z)=&-n\sum_{m=1}^{\infty}m\kappa_{-n,m}z^{-m-1},\\
\frac{(g^{-1})'(z)}{g^{-1}(z)}=&\frac{1}{z}+ \sum_{m=1}^{\infty} m\kappa_{0,m}z^{-m-1}.
\end{split}
\end{equation*}
In a neighbourhood of $z=0$,
\begin{equation*}
\begin{split}
\mathcal{P}_n' \circ f^{-1}(z)(f^{-1})'(z)=& n\sum_{m=1}^{\infty} m\kappa_{n,-m}z^{m-1},\\
\mathcal{Q}_n' \circ f^{-1}(z)(f^{-1})'(z)=&-nz^{-n-1}+n\sum_{m=1}^{\infty}m\kappa_{-n,-m}z^{m-1},\\
\frac{(f^{-1})'(z)}{f^{-1}(z)}=& \frac{1}{z}-\sum_{m=1}^{\infty} m\kappa_{0,-m}z^{m-1}.
\end{split}
\end{equation*}
The definitions \eqref{eq4_12_6} and \eqref{eq4_12_7} of $t_n, n\in \Z,$ and Lemma \ref{l4_12_1} then shows that $\pa_n t_m =\delta_{n,m}$.
\end{proof}

As follows from this proposition, $\{t_n, n\in\Z\}$ are good local coordinates on $\mathfrak{D}$. Therefore the partial derivatives $\frac{\pa}{\pa t_n}$ and $\frac{\pa}{\pa \bar{t}_n}$ are well defined and $\frac{\pa}{\pa t_n}$ coincides with the  variation $\pa_n$ defined in the proposition above. We then have

\begin{proposition}\label{p4_12_2}
The variation of the functions $v_m, m\in\Z,$ with
respect to the coordinates $t_n$, $n \in \Z,$ is given by the
following. For $m\neq 0$,
\begin{equation}\label{eq4_12_10}\begin{split}
\frac{\pa v_m}{\pa t_n} = -|mn|\kappa_{n,m} , \hspace{0.2cm}
n\neq 0,\hspace{1cm}\frac{\pa v_m}{\pa t_0} = |m|\kappa_{0,m},\end{split}
\end{equation}
and for $m=0$,
\begin{equation}\label{eq4_12_9}
\frac{\pa v_0}{\pa t_n}= |n|\kappa_{n,0},\hspace{0.2cm} n\neq
0,\hspace{1cm} \frac{\pa v_0}{\pa t_0} = -2\kappa_{0,0}.
\end{equation}
Moreover,
\begin{equation}\label{eq4_12_8}
\frac{\pa v_m}{\pa \bar{t}_n}=0, \hspace{1cm}\text{for all}\;
m,n.
\end{equation}
\end{proposition}
\begin{proof}\eqref{eq4_12_8} follows immediately since both $t_n, n\in\Z,$ and $v_m, m\in \Z,$ depend on the coefficients of $f$ and $g$ holomorphically. \eqref{eq4_12_10} follows from the definition \eqref{eq4_12_6} of $v_m$ and the proof of Proposition \ref{p4_12_1}. For \eqref{eq4_12_9},
a straightforward computation give
\begin{equation*}
\begin{split}
&\frac{\pa}{\pa t_n}\left\{ \log\frac{g(w)}{w}
\frac{ g'(w)}{f(w)}-  \log
\frac{f(w)}{w} g(w)
\frac{f'(w)}{f(w)^2}- \frac{1}{w}\frac{g(w)}{f(w)}\right\}\\
=&\frac{1}{f(w)} \frac{\pa g(w)}{\pa t_n}\frac{ g'(w)}{g(w)}+\frac{1}{f(w)}\log\frac{g(w)}{w}\frac{\pa}{\pa w}\left(\frac{\pa g(w)}{\pa t_n}\right)-\log\frac{g(w)}{w}
\frac{ g'(w)}{f^2(w)}\frac{\pa f(w)}{\pa t_n}\\&-   g(w)
\frac{f'(w)}{f(w)^3}\frac{\pa f(w)}{\pa t_n}- \log
\frac{f(w)}{w} \frac{\pa g(w)}{\pa t_n}
\frac{f'(w)}{f(w)^2}-  \log
\frac{f(w)}{w} g(w)\frac{\pa}{\pa w}\left(\frac{1}
{f(w)^2} \frac{\pa f(w)}{\pa t_n}\right)\\
&-\frac{1}{wf(w)}\frac{\pa g(w)}{\pa t_n}+\frac{g(w)}{w f(w)^2}\frac{\pa f(w)}{\pa t_n}\\
=&\frac{\pa}{\pa w}\left( \frac{1}{f(w)} \frac{\pa g(w)}{\pa t_n}\log\frac{g(w)}{w}-  \log
\frac{f(w)}{w} \frac{ g(w)}
{f(w)^2} \frac{\pa f(w)}{\pa t_n} \right)\\&+\frac{1}{f(w)^2}\left(\log\frac{f(w)}{w}-\log\frac{g(w)}{w}\right)\left(g'(w)\frac{\pa f(w)}{\pa t_n}-f'(w)\frac{\pa g(w)}{\pa t_n}\right).
\end{split}
\end{equation*}The definition \eqref{eq4_12_7} of $v_0$ then implies that
\begin{equation*}
\frac{\pa v_0}{\pa t_n}=\frac{1}{2\pi i}\oint_{S^1} \frac{1}{f(w)^2}\left(\log\frac{f(w)}{w}-\log\frac{g(w)}{w}\right)\left(g'(w)\frac{\pa f(w)}{\pa t_n}-f'(w)\frac{\pa g(w)}{\pa t_n}\right)dw.
\end{equation*}We obtain from the proof of Proposition \ref{p4_12_1} that
\begin{equation*}
\frac{\pa v_0}{\pa t_0} =-\frac{1}{2\pi i}\oint_{S^1}\frac{1}{w}\left(\log\frac{f(w)}{w}-\log\frac{g(w)}{w}\right)dw=-\log a_1+\log b=2\log b=-2\kappa_{0,0},
\end{equation*}and for $n\geq 1$,
\begin{equation*}
\begin{split}
\frac{\pa v_0}{\pa t_n}=&-\frac{1}{2\pi i }\oint_{S^1}\mathcal{P}_n'(w) \left(\log\frac{f(w)}{w}-\log\frac{g(w)}{w}\right)dw=\frac{1}{2\pi i}\oint_{S^1} \mathcal{P}_n'(w) \log \frac{g(w)}{w}dw\\
=&-\frac{1}{2\pi i}\oint_{\mathcal{C}_2} \mathcal{P}_n'\circ g (z) (g^{-1})'(z) \log \frac{g^{-1}(z)}{z}dz=n\kappa_{n,0},\\
\frac{\pa v_0}{\pa t_{-n}}=&-\frac{1}{2\pi i }\oint_{S^1}\mathcal{Q}_n'(w) \left(\log\frac{f(w)}{w}-\log\frac{g(w)}{w}\right)dw=-\frac{1}{2\pi i}\oint_{S^1} \mathcal{Q}_n'(w) \log \frac{f(w)}{w}dw\\
=&\frac{1}{2\pi i}\oint_{\mathcal{C}_1} \mathcal{Q}_n'\circ f (z) (f^{-1})'(z) \log \frac{f^{-1}(z)}{z}dz=n\kappa_{-n,0}.\\
\end{split}
\end{equation*}
\end{proof}

Since the Grunsky coefficients $\kappa_{m,n}$ are symmetric,    Proposition
\ref{p4_12_2} shows that there should formally
exist a function $\mathcal{F}$ on $\mathfrak{D}$ such that $\pa
\F/\pa t_n = v_n$.
Let $\Psi(z)$ be a holomorphic function on $\Omega_1^+$ which has a series expansion
\begin{equation*}
\Psi(z) =\sum_{n=1}^{\infty} \frac{v_{-n}}{n}z^n
\end{equation*}in a neighbourhood of $z=0$; and let $\Phi(z)$ be a meromorphic function on $\Omega_2^-$ which has a series expansion
\begin{equation*}\Phi(z)=\sum_{n=1}^{\infty}\frac{v_n}{n}z^{-n}
\end{equation*} in a neighbourhood of $z=\infty$. Notice that $\Psi'(z)= -\tilde{S}_+(z)$ and
$\Phi'(z)=S_{-}(z)+t_0/z$. From Proposition \ref{p4_12_2} and \eqref{iden1}, we have
\begin{lemma} The
variations of the functions $\Phi$ and $\Psi$ with respect to $t_n, n\in \Z$ are given by
\begin{align*}
\frac{\pa \Psi}{\pa t_n}(z)&=-\mathcal{P}_n(f^{-1}(z)) + n
\kappa_{n,0}, \hspace{1.3cm}
\frac{\pa \Phi}{\pa t_n}(z) =-\mathcal{P}_n(g^{-1}(z))+ z^n,\\
\frac{\pa \Psi}{\pa t_0}(z)&= -\log \frac{f^{-1}(z)}{z}-\log
a_1,\hspace{1.5cm}\frac{\pa
\Phi}{\pa t_0}(z) = -\log \frac{g^{-1}(z)}{z}-\log b,\\
\frac{\pa \Psi}{\pa t_{-n}}(z)&= -\mathcal{Q}_n(f^{-1}(z))+
z^{-n},\hspace{1.5cm}\frac{\pa \Phi}{\pa
t_{-n}}(z)=-\mathcal{Q}_n(g^{-1}(z))-n\kappa_{-n,0}.
\end{align*}
\end{lemma}

 Now define the tau function $\tau$ by
 \begin{equation}\label{eq5_1_1}
 \tau =|\mathfrak{T}|^2=\mathfrak{T}\bar{\mathfrak{T}},
 \end{equation}where $\mathfrak{T}$ is a holomorphic function on $\mathfrak{D}$ defined by
\begin{equation}\label{eq4_12_11}\begin{split}
 \log \mathfrak{T} =  &\frac{t_0 v_0}{2} -\frac{t_0^2}{4} +\frac{1}{8\pi i}\oint_{S^1}
\frac{g'(w)}{f(w)}\Bigl\{g(w)\Phi'(g(w))+
2\Phi(g(w))\Bigr\} dw \\
&+\frac{1}{8\pi i}\oint_{S^1}
\frac{g(w)f'(w)}{f(w)^2}\Bigl\{f(w)\Psi'(f(w)) -
2\Psi(f(w))\Bigr\} dw.\end{split}
\end{equation}When the sum converges absolutely, $\log \mathfrak{T}$ can be written
explicitly as
\begin{align*}
\log \mathfrak{T} = \frac{t_0 v_0}{2} -\frac{t_0^2}{4} -\frac{1}{4}\sum_{n=1}^{\infty}(n-2)(t_n
v_n+t_{-n}v_{-n}).
\end{align*}By  definition, $\log\tau$ is a harmonic function on $\mathfrak{D}$.
 The partial derivatives of $\log\tau$ with respect to $t_n$ are given by the following proposition:
\begin{proposition}\label{p4_27_1}For all $n\in\Z$, we have
\begin{align*}
\frac{\pa \log \tau}{\pa t_n}=v_n, \hspace{3cm}\frac{\pa \log
\tau}{\pa \bar{t}_n}=\bar{v}_n.
\end{align*}
\end{proposition}
\begin{proof}Since $\log\tau =\log\mathfrak{T}+\overline{\log\mathfrak{T}}$ and $\mathfrak{T}$ is holomorphic, it is sufficient to prove that $$\frac{\pa \log\mathfrak{T}}{\pa t_n}=v_n.$$
We consider only the case $n\geq 1$. The case where $n\leq 0$ is similar. From \eqref{eq4_12_11}, we have
\begin{equation}\label{eq4_12_12}\begin{split}
4\frac{\pa\log \mathfrak{T}}{\pa t_n} =  2t_0 \frac{\pa v_0}{\pa t_n} &+\frac{1}{2\pi i}\oint_{\mathcal{C}_2}
\left(\frac{\pa}{\pa t_n}\frac{1}{f\circ g^{-1}(z)}\right)\Bigl\{z\Phi'(z)+
2\Phi(z)\Bigr\} dz\\&+\frac{1}{2\pi i}\oint_{\mathcal{C}_2}
 \frac{1}{f\circ g^{-1}(z)}  \left\{z\frac{\pa}{\pa z}\frac{\pa \Phi(z)}{\pa t_n}+
2\frac{\pa\Phi(z)}{\pa t_n}\right\} dz\\
&+\frac{1}{2\pi i}\oint_{\mathcal{C}_1}
\frac{1}{z^2}\frac{\pa (g\circ f^{-1})(z)}{\pa t_n}\Bigl\{z\Psi'(z) -
2\Psi(z)\Bigr\} dz\\
&+\frac{1}{2\pi i}\oint_{\mathcal{C}_1}
\frac{g\circ f^{-1}(z)}{z^2} \left\{z\frac{\pa}{\pa z}\frac{\pa \Psi(z)}{\pa t_n} -
2\frac{\pa\Psi(z)}{\pa t_n}\right\} dz.  \end{split}
\end{equation}Now $\frac{\pa v_0}{\pa t_n}=n\kappa_{n,0}$,
\begin{equation*}
\begin{split}
&\frac{1}{2\pi i}\oint_{\mathcal{C}_2}
\left(\frac{\pa}{\pa t_n}\frac{1}{f\circ g^{-1}(z)}\right)\Bigl\{z\Phi'(z)+
2\Phi(z)\Bigr\} dz\hspace{4cm}\\=&\frac{1}{2\pi i}\oint_{\mathcal{C}_2}
 \frac{\pa S_+(z)}{\pa t_n} \Bigl\{z\Phi'(z)+
2\Phi(z)\Bigr\} dz=(-n+2)v_n,
\end{split}
\end{equation*}
\begin{equation*}
\begin{split}
&\frac{1}{2\pi i}\oint_{\mathcal{C}_2}
 \frac{1}{f\circ g^{-1}(z)}  \left\{z\frac{\pa}{\pa z}\frac{\pa \Phi(z)}{\pa t_n}+
2\frac{\pa\Phi(z)}{\pa t_n}\right\} dz\\=&\frac{1}{2\pi i}\oint_{\mathcal{C}_2} \frac{1}{f\circ g^{-1}(z)} \Bigl\{(n+2)z^n -z\mathcal{P}_n'(g^{-1}(z))(g^{-1})'(z)-2\mathcal{P}_n(g^{-1}(z))\Bigr\}dz\\
=&\frac{1}{2\pi i}\oint_{S^1}\frac{1}{f(w)} \Bigl\{(n+2)g(w)^n g'(w)-g(w)\mathcal{P}_n'(w)-2\mathcal{P}_n(w)g'(w)\Bigr\}dw\\
=&(n+2)v_n -\frac{1}{2\pi i} \oint_{S^1}\frac{1}{f(w)} \Bigl\{ g(w)\mathcal{P}_n'(w)+2\mathcal{P}_n(w)g'(w)\Bigr\}dw,
\end{split}
\end{equation*}
\begin{equation*}
\begin{split}
&\frac{1}{2\pi i}\oint_{\mathcal{C}_1}
\frac{1}{z^2}\frac{\pa (g\circ f^{-1})(z)}{\pa t_n}\Bigl\{z\Psi'(z) -
2\Psi(z)\Bigr\} dz
\hspace{4.3cm}\\=&\frac{1}{2\pi i}\oint_{\mathcal{C}_1} \frac{\pa\tilde{S}_-(z)}{\pa t_n} \Bigl\{z\Psi'(z) -
2\Psi(z)\Bigr\} dz=0,
\end{split}
\end{equation*}
\begin{equation*}
\begin{split}
&\frac{1}{2\pi i}\oint_{\mathcal{C}_1}
\frac{g\circ f^{-1}(z)}{z^2} \left\{z\frac{\pa}{\pa z}\frac{\pa \Psi(z)}{\pa t_n} -
2\frac{\pa\Psi(z)}{\pa t_n}\right\} dz \hspace{4cm}\\
=&-2nt_0\kappa_{n,0} -\frac{1}{2\pi i}\oint_{S^1}\frac{g(w)}{f(w)^2}\Bigl\{f(w)\mathcal{P}_n'(w)-2\mathcal{P}_n(w)f'(w)\Bigr\}dw.
\end{split}
\end{equation*}Adding the terms, we find that
\begin{equation*}
4\frac{\pa\log\mathfrak{T}}{\pa t_n}= 4v_n -\frac{1}{\pi i}\oint_{S^1} \frac{d}{dw} \frac{g(w)\mathcal{P}_n(w)}{f(w)}dw=4v_n,
\end{equation*}which is the desired assertion.
\end{proof}

Using this proposition and Proposition \eqref{p4_12_2}, we find that
\begin{corollary}For the function $\tau$ defined by \eqref{eq4_12_11}, we have
\begin{equation*}
\begin{split}\frac{\pa^2\log\tau}{\pa t_m \pa t_n} =\begin{cases}
-|mn|\kappa_{m,n}, \hspace{0.5cm}&\text{if}\;\; m\neq 0, n\neq 0,\\
|m|\kappa_{m,0}, & \text{if} \;\; m\neq 0, n=0,\\
-2\kappa_{0,0}, &\text{if}\;\; m=n=0,
\end{cases}
\end{split}
\end{equation*}where $\kappa_{m,n}$ is the generalized Grunsky coefficients of the pair of univalent functions $(f^{-1}, g^{-1})$.
\end{corollary}
It follows from Proposition \ref{Hirota} that \begin{theorem}\label{th1}
The evolutions of the pair of conformal mappings
$( g, f)$ with respect to $t_n, n\in \Z,$ satisfy the dispersionless Toda
hierarchy \eqref{Lax}.\end{theorem}

Since  the coefficients of $f
$ and $g$ depend holomorphically on the variables $t_n$, $n\in\Z$, and $\log\tau$ is a harmonic function on $\mathfrak{D}$, it follows that  we have another solution of  the dispersionless Toda
hierarchy if we take $\bar{t}_n$ as the time variables. More precisely, let $\bar{f}$ and $\bar{g}$ be conformal mappings defined respectively by $\bar{f}(z)=\overline{f(\bar{z})}$ and $\bar{g}(z)=\overline{g(\bar{z})}$. Then  \begin{theorem}
The evolutions of the pair of conformal mappings
$( \bar{g}, \bar{f})$ with respect to $\bar{t}_n, n\in\Z,$ satisfy the dispersionless Toda
hierarchy \eqref{Lax}.\end{theorem}

An interesting thing to note is that on the subspace of $\mathfrak{D}$ defined by $f=\bar{f}$ and $g=\bar{g}$, the coefficients of $f$ and $g$ are real. Therefore $t_n$ are real variables. It follows that restricted to this totally real submanifold of $\mathfrak{D}$, we have the usual dispersionless Toda flows where the time variables $t_n, n\in\Z,$ are real.

\section{Symplectic structure and Riemann Hilbert data}

In the language of Takasaki and Takebe (see \cite{TT1} and the
references therein), to every solution of the dispersionless Toda
hierarchy, one can associate a Riemann Hilbert data (or called the
twistor data). Namely there exist two pairs of functions $(r,h)$
and $(\tilde{r}, \tilde{h})$ of the variables $w$ and $t_0$ such
that
\begin{align}\label{RH}
\{r, h\}_T &= r, \hspace{3cm} \{\tilde{r}, \tilde{h}\}_T=
\tilde{r} , \\
r(\mL, \mM) &= \tilde{r}(\tilde{\mL}, \tilde{\mM}), \hspace{1.5cm}
h(\mL, \mM) = \tilde{h}(\tilde{\mL}, \tilde{\mM}).\nonumber
\end{align}
Here $\mM$ and $\tilde{\mM}$ are the Orlov-Schulman functions.
They are defined so that they can be written as
\begin{equation}\label{orlov}\begin{split}
\mM&= \sum_{n=1}^{\infty} nt_n \mL^{n} + t_0 + \sum_{n=1}^{\infty}
v_n \mL^{-n} ,\\
\tilde{\mM} &=-\sum_{n=1}^{\infty} nt_{-n} \tilde{\mL}^{-n} + t_0
- \sum_{n=1}^{\infty} v_{-n} \tilde{\mL}^n;\nonumber\end{split}
\end{equation}
and they form a canonical pair with $\mL$ and $\tilde{\mL}$, i.e.
\begin{equation}\label{eq4_28_3}\begin{split}\{\mL, \mM\}_T=w\frac{\pa \mL}{\pa w}\frac{\pa \mM}{\pa t_0} - w\frac{\pa \mM}{\pa w}\frac{\pa \mL}{\pa t_0}=\mL\\  \{\tilde{\mL},
\tilde{\mM}\}_T=w\frac{\pa \tilde{\mL}}{\pa w}\frac{\pa \tilde{\mM}}{\pa t_0} - w\frac{\pa \tilde{\mM}}{\pa w}\frac{\pa \tilde{\mL}}{\pa t_0}=\tilde{\mL}.\end{split}\end{equation} Conversely, Takasaki and Takebe also
showed that if $(\mL, \tilde{\mL})$ are formal power series of the
form \eqref{series}, $\mM ,\tilde{\mM}$ are formal functions of
the form \eqref{orlov}, and there exist $(r, h)$ , $(\tilde{r},
\tilde{h})$ satisfying \eqref{RH}, then $(\mL, \tilde{\mL})$ is a
solution to the dispersionless Toda hierarchy. The proof is
formal. The main technique is comparing powers of $w$. However
nothing is assumed about the convergence of the series. For the
 conformal mapping problems we  study in the previous section,   $\mL=g,
\tilde{\mL}=f$. If we define $\mM$ and $\tilde{\mM}$ as \eqref{orlov},  then compare to \eqref{eq4_28_1} and \eqref{eq4_28_2}, we find that restricted to the unit circle $S^1$,
\begin{equation}\label{eq4_28_4}\begin{split}
\mM(w) =g(w)\left( S_+(g(w))-S_-(g(w))\right) = \frac{g(w)}{f(w)},\\
\tilde{\mM}(w) = f(w) \left(\tilde{S}_+(f(w))-\tilde{S}_-(f(w))\right)=\frac{g(w)}{f(w)}.
\end{split}
\end{equation}In other words, \begin{equation}\label{eq4_28_5}\mM=\tilde{\mM}.\end{equation} Moreover, \begin{equation}\label{eq4_28_6}\tilde{\mL}\mM=\mL \hspace{0.5cm}\text{or}\hspace{0.5cm} \tilde{\mL}\tilde{\mM}=\mL.\end{equation} To prove the identities \eqref{eq4_28_3} in our context, notice that \eqref{orlov} shows that for $w\in S^1$,
\begin{equation}\begin{split}
\frac{1}{\mL}\left\{ \mathcal{L}, \mathcal{M}\right\}_T = &\frac{1}{g(w)}\left\{ g(w), t_0\right\}_T +\sum_{n=1}^{\infty} \left\{g(w), v_n\right\}_Tg(w)^{-n-1}\\
=&w\left\{ \frac{g'(w) }{g(w)}+\sum_{n=1}^{\infty} g'(w) \frac{\pa v_n}{\pa t_0} g(w)^{-n-1}\right\}.
\end{split}
\end{equation}Using \eqref{eq4_12_10} and \eqref{iden1}, this gives
\begin{equation*}
\begin{split}
\frac{1}{\mL}\left\{ \mathcal{L}, \mathcal{M}\right\}_T=&w\frac{\pa}{\pa w}\left\{ \log g(w) - \sum_{n=1}^{\infty} \kappa_{n,0} g(w)^{-n}\right\}\\
=&w\frac{\pa}{\pa w}\left\{ \log g(w) +\log \frac{g^{-1}(g(w))}{g(w)}-\log b^{-1}\right\}=1.
\end{split}
\end{equation*}The identity $\left\{\tilde{\mL}, \tilde{\mM}\right\}_T=\tilde{\mL}$ can be proved in a similar way, or by observing that \eqref{eq4_28_5} and \eqref{eq4_28_6} give
\begin{equation*}
\frac{1}{\tilde{\mL}}\left\{\tilde{\mL},\tilde{\mM}\right\}_T = \frac{1}{\tilde{\mL}\tilde{\mM}}\left\{ \tilde{\mL}\tilde{\mM},\tilde{\mM}\right\}_T=\frac{1}{\mL}\left\{ \mathcal{L}, \mathcal{M}\right\}_T=1.
\end{equation*}It is interesting to note that the identity $\frac{1}{\mL}\left\{ \mathcal{L}, \mathcal{M}\right\}_T=1$ can also be written as
\begin{equation*}
1= \left\{ \mathcal{L}, \mathcal{L}^{-1}\mathcal{M}\right\}_T=\left\{\mL, \tilde{\mL}^{-1}\right\}_T=\left\{ g(w), f(w)^{-1}\right\}_T,
\end{equation*}which is known as the string equation.  We can read from \eqref{eq4_28_5} and \eqref{eq4_28_6} that the Riemann-Hilbert data for our problem is \begin{equation}\label{eq4_29_3}\begin{split}r(w, t_0)= w, \hspace{1cm}&\tilde{r}(w, t_0)=pt_0,\\h(w, t_0)=t_0, \hspace{1cm}&\tilde{h}(w, t_0)=t_0.\end{split}\end{equation} On the other hand, we can transform the pair of equations \eqref{eq4_28_5} and \eqref{eq4_28_6} to
\begin{equation*}\mL=\tilde{\mL}\tilde{\mM},\hspace{1cm} \tilde{\mL}^{-1}=\mathcal{L}^{-1}\mM,
\end{equation*}which is the Riemann Hilbert problem
studied by Takasaki \cite{TT3} in connection to string theory.

\section{Integrable structure of comformal mappings $\grave{\text{a}}$ la Wiegmann-Zabrodin}\label{s1}
In this section, we would like to discuss the relation between the integrable structure on pairs of conformal mappings discussed in Section \ref{ext} and  the integrable structure  of conformal mappings
observed by Wiegmann and Zabrodin \cite{WZ, KKMWZ, MWZ}. In the
approach of Wiegman and Zabrodin, the coordinates $t_n$ are
defined as harmonic moments. Given a domain $\Omega^-$ containing $\infty$, bounded by
an analytic curve $\mathcal{C}$ and with complement $\Omega^+$, define
\begin{equation}\label{eq4_28_9}\begin{split}
t_n &=\frac{1}{2 \pi i}\oint_{\mathcal{C}} z^{-n} \z dz,
\hspace{1cm} v_n =\frac{1}{2\pi i} \oint_{\mC} z^n \z dz,\hspace{1cm}n\geq 1;\\
t_0&=\frac{1}{2\pi i} \oint_{\mC} \z dz , \hspace{1.8cm}
v_0=\frac{1}{\pi }\iint\limits_{\Omega^+} \log |z|^2 d^2z.
\end{split}\end{equation}
For $n\geq 1$, let $t_{-n}=-\bar{t}_n$ and  $v_{-n}=-\bar{v}_n$. Denote by
$g$  the conformal map mapping the exterior disc to the
domain $\Omega^-$ normalized so that $g(\infty)=\infty$,
$g'(\infty) >0$. Let
\begin{align}\label{sol}
\mL(w)=g(w),\hspace{1cm}\text{and}\hspace{1cm}
\tilde{\mL}(w)=\frac{1}{\overline{g(1/\bar{w})}}.
 \end{align}
 Wiegmann and Zabrodin show
that $(\mL, \tilde{\mL})$ is a solution of the  dispersionless
Toda hierarchy with respect to the time variables $t_n, n\in \Z$.

Consider the transformation on the space $\mathfrak{D}$ defined by \begin{equation}\label{eq4_28_8}
(f,g)\mapsto \left( \tilde{f}(w) =\frac{1}{\overline{g(1/\bar{w})}}, \tilde{g}(w) =\frac{1}{\overline{f(1/\bar{w})}} \right).
\end{equation}Under this transformation, it is easy to check that\begin{equation*}\begin{split}
\tilde{t}_n =-\bar{t}_{-n} \;\;\;\;\text{for all}\;\; n\neq 0,\hspace{0.5cm}\text{and}\hspace{0.5cm} \tilde{t}_0=\bar{t}_0,\\\tilde{v}_n =-\bar{v}_{-n} \;\;\;\;\text{for all}\;\; n\neq 0,\hspace{0.5cm}\text{and}\hspace{0.5cm} \tilde{v}_0=\bar{v}_0.\end{split}
\end{equation*}In other words, in terms of the coordinates $t_n, n\in \Z$, the transformation \eqref{eq4_28_8} is furnished by the automorphism $$\left\{t_n\right\} \mapsto \left\{ t_n'\right\}, \hspace{1cm}\text{where}\;\; t_n'=\begin{cases} -\bar{t}_{-n}, \hspace{0.5cm}&\text{if}\;\; n\neq 0,\\
\bar{t}_0, \hspace{0.5cm} &\text{if} \;\; n=0, \end{cases}$$ of $\mathfrak{D}$. The invariant subspace of this automorphism is the space $\Sigma$ defined by the equations $\bar{t}_n =-t_{-n}$ for $n\neq 0$ and $t_0=\bar{t}_0$, which is the space containing all pairs of conformal mappings of the form $(1/\overline{g(1/\bar{w})}, g(w))$. It is straightforward to check that on the subspace $\Sigma$, the definitions of the variables $t_n$ and $v_n$ \eqref{eq4_12_6} reduce to \eqref{eq4_28_9}. Notice that $\Sigma$ is not a complex manifold because of the one extra dimension furnished by the real variable $t_0$. We can take the variables $ t_0, \text{Re}\, t_n, \text{Im}\,t_n$, $n\geq 1$ as coordinates on the real manifold $\Sigma$. To make some distinctions, we denote by $\mathrm{t}_n$ the variables $t_n$ restricted to $\Sigma$ so that for $n\neq 1$, $\mathrm{t}_{-n}=-\bar{\mathrm{t}}_n$. Any function $\mathcal{F}_{\mathfrak{D}}(t_n, \bar{t}_n)$ on $\mathfrak{D}$ restricted to the function $\mathcal{F}_{\Sigma}\left( \mathrm{t}_n\right)=\mathcal{F}_{\mathfrak{D}}\left(\mathrm{t}_n, -\mathrm{t}_{-n}\right) $ on $\Sigma$. Therefore, the partial derivatives $\frac{\pa}{\pa \mathrm{t}_n}$ on $\Sigma$ can be defined in terms of the partial derivatives $\frac{\pa }{\pa t_n}$ and $\frac{\pa}{\pa \bar{t}_n}$ by
\begin{equation*}
\frac{\pa}{\pa \mathrm{t}_n}=\frac{\pa }{\pa t_n}-\frac{\pa}{\pa \bar{t}_{-n}}, \hspace{0.5cm}\text{for}\;\; n\neq 0,\hspace{0.5cm}\text{and}\hspace{0.5cm}
\frac{\pa}{\pa \mathrm{t}_0}=\frac{\pa }{\pa t_0}+\frac{\pa}{\pa \bar{t}_{0}}.
\end{equation*}Notice that $\frac{\pa}{\pa \mathrm{t}_n}$ are well-defined (complex) vector fields on $\Sigma$ since they annihilate the defining functions $Z_n(t_m, \bar{t}_m)=\bar{t}_n+t_{-n}, n\in\Z,$ of $\Sigma$. Now since the functions $\mathfrak{T}$ \eqref{eq4_12_11} and $v_n$ \eqref{eq4_12_6} on $\mathfrak{D}$ are holomorphic, we find immediately from the results in Section \ref{ext} that their restrictions to $\Sigma$ satisfies
\begin{equation}\label{eq4_29_1}
\begin{split}
\frac{\pa \log \mathfrak{T}_{\Sigma}}{\pa \mathrm{t}_n}=v_n, \hspace{1cm}\frac{\pa^2\log \mathfrak{T}_{\Sigma}}{\pa \mathrm{t}_m\pa \mathrm{t}_n}=
\begin{cases}
-|mn|\kappa_{m,n},\hspace{0.5cm}&\text{if}\;\; mn\neq 0\\
|m|\kappa_{m,0}, &\text{if}\;\;m\neq 0, n=0,\\
-2\kappa_{0,0}, &\text{if} \;\; m=n=0.
\end{cases}
\end{split}
\end{equation} Proposition \ref{Hirota} then shows that $(1/\overline{g(1/\bar{w})}, g(w))$ is a solution of the dispersionless Toda hierarchy with respect to the time variables $\mathrm{t}_n, n\in \Z$, with   restriction $\mathrm{t}_{-n}=-\bar{\mathrm{t}}_n$. This is precisely the result of Wiegmann and Zabrodin \cite{WZ, KKMWZ, MWZ}. The corresponding tau function is the restriction of $\mathfrak{T}$ \eqref{eq4_12_11} to $\Sigma$. We left it as an exercise for the reader to show that restricted to $\Sigma$, the function $\mathfrak{T}$ is given by
\begin{equation*}
\log\mathfrak{T}_{\Sigma}= -\frac{1}{\pi^2}\int\limits_{\Omega^-}\int\limits_{\Omega^-}\log\left|\frac{1}{z}-\frac{1}{\zeta}\right|d^2\zeta d^2z,
\end{equation*}which is a real-valued function. Therefore, restricted to $\Sigma$, $\tau =\mathfrak{T}^2$. As a result, the restriction of the tau function $\tau$ on $\mathfrak{D}$ to $\Sigma$ is not   the corresponding tau function on $\Sigma$.

\section{Conformal weldings and dispersionless Toda hierarchy}
In this section, we review the concept of conformal weldings and discuss their evolutions under the dispersionless Toda flow. For details about conformal weldings, one can see \cite{Ki, KY, Lehto, Gar}.

Let $\Homeo_{C}(S^1)$ be the space of all $C^1$
homeomorphisms on the unit circle $S^1$.  Notice that a
$C^1$ homeomorphism $\gamma \in \Homeo_{C}(S^1)$ is also a quasi-symmetric homeomorphism, i.e., $\gamma(e^{i\theta})$ satisfies the inequality
\begin{align}\label{def1}\frac{1}{M}\leq \frac{\gamma(e^{
i(\theta+ \omega)})-\gamma(e^{ i \theta})}{\gamma(e^{  i\theta})-\gamma(e^{
i (\theta-\omega)})} \leq M, \hspace{1cm}\forall \;\;\theta, \omega\in \R,
0<\omega<\frac{\pi}{2},\end{align} for some constant $M>1$. Therefore, according to the theory of quasiconformal mappings,
$\gamma$ can be extended to be a $C^1$ map on the
extended complex plane $\hat{\C}=\C\cup\{\infty\}$, which is also denoted by $\gamma$, and satisfies
\begin{align*}
\gamma\left(\frac{1}{\z}\right)=\frac{1}{\ov{\gamma(z)}},
\hspace{2cm}\forall z\in \C.
\end{align*}
Moreover, $\gamma$ is real analytic on $\hat{\C} \setminus S^1$.

In \cite{Teo04}, we  used the theory of quasiconformal mappings  to show that given a quasi-symmetric homeomorphism with its quasiconformal extension $\gamma$, there exist quasiconformal mappings $\tilde{f}$ and $\tilde{g}$ such that $\gamma= \tilde{g}^{-1}\circ \tilde{f}$, and $\tilde{f}\bigr\vert_{\mathbb{D}}$ and $\tilde{g}\bigr\vert_{\mathbb{D}^*}$ are univalent functions. Moreover, $\tilde{f}$ and $\tilde{g}$ are unique if we impose the conditions $\tilde{f}(0)=0$, $\tilde{f}'(0)=1$ and $\tilde{g}(\infty)=\infty$. Define $f=r\circ \tilde{f}$ and $g=r\circ \tilde{g}$, where $r$ is a complex number so that $r^2 =1/\tilde{g}'(\infty)$, we find that $\gamma=g^{-1}\circ f$ and $f'(0)g'(\infty) = r^2\tilde{f}'(0)\tilde{g}'(\infty)=1$. In other words, we have shown that given $\gamma \in \Homeo_{C}(S^1)$, there exists two $C^1$  homeomorphisms $f$ and $g$ of the plane, such
that $\gamma= g^{-1}\circ f$, and $f\bigr\vert_{\mathbb{D}}$ and
$g\bigr\vert_{\mathbb{D}^*}$ are the unique univalent functions
satisfying $f(0)=0$, $g(\infty)= \infty$ and
$f'(0)g'(\infty)=1$. The decomposition of $\gamma$ as
$g^{-1}\circ f$ is known as conformal welding or sewing\footnote{The conformal welding of $\gamma\in S^1\bk\text{Diff}_+(S^1)$, where $\text{Diff}_+(S^1)$ is the space of diffeomorphisms on the unit circle, was first discussed in \cite{Ki}.}.

Given $\gamma \in \Homeo_{C}(S^1)$ with conformal welding $\gamma=g^{-1}\circ f$,
we can associate $\gamma$ with the simply connected domain $\Omega^+=f(\mathbb{D}) =
g(\mathbb{D})$, its exterior $\Omega^-=f(\Del^*)=g(\Del^*)$ and their common boundary
$\mathcal{C} = f(S^1)=g(S^1)$, a $C^1$ curve. However, such an association
is not one-to-one. If $\gamma_1  =g_1^{-1}\circ f_1$ and $\gamma_2=g_2^{-1}\circ f_2$ are associated to the same domain, then $f_1(\Del) =f_2(\Del)$ implies that $f_1^{-1}\circ f_2$ is a univalent function on $\Del$ mapping the unit disc back to itself. Therefore, $f_1^{-1}\circ f_2$ is a linear fractional transformation of the form $e^{i\theta}\frac{z+a}{1+\bar{a} z}$ for some $a\in \Del$ and $\theta\in \R$. However, the condition $f_1(0)=f_2(0)=0$ forces $a=0$. Therefore, we are left with the possibility $f_2(z)=f_1(e^{i\theta }z)$. Similar argument shows that $g_1(\Del^*)=g_2(\Del^*)$ implies that $g_2(z)=g_1(e^{i\omega}z)$ for some $\omega\in\R$. The condition $f_j'(0)g_j'(\infty)=1$, $j=1,2$, then force $e^{i\theta}=e^{i\omega}$. On the other hand, it is easy to show that given $\gamma\in \Homeo_{C}(S^1)$ with conformal welding $\gamma=g^{-1}\circ f$ and given $r\in S^1$, the conformal welding of $r^{-1}\circ \gamma\circ r\in \Homeo_{C}(S^1)$  is given by $\gamma=(g\circ r)^{-1}\circ (f\circ r)$ and therefore $\gamma$ and $r^{-1}\circ \gamma \circ r$ are both associated to the domain $\Omega^+=f(\Del)=f\circ r(\Del)$. As a conclusion, $\gamma_1$ and $\gamma_2$ are associated with the same domain $\Omega^+$ if and only if $\gamma_2=r^{-1}\circ \gamma_1\circ r$ for some $r\in S^1$.

Now return to our discussion on the evolutions of conformal mappings, we see from the unique decomposition $\text{Homeo}_{C}(S^1)\ni \gamma =g^{-1}\circ f $ that we can identify $\text{Homeo}_{C}(S^1)$ as a subspace of $\mathfrak{D}$ containing the pairs $(f,g)$ with $f(S^1)=g(S^1)$. Unlike the subspace $\Sigma$ which can be easily identified as the subspace of $\mathfrak{D}$ defined by $\bar{t}_n =-t_{-n}$, the characterization of the space $\text{Homeo}_{C}(S^1)$ is a highly nontrivial issue. Assume that the subspace $\text{Homeo}_{C}(S^1)$ can be defined locally by $\bar{t}_n = \mathcal{Z}_n ( t_m)$. Then we can take $\mathrm{t}_n =t_n$ as a set of local parameters on $\text{Homeo}_{C}(S^1)$ so that  any function $\mathcal{F}_{\mathfrak{D}}(t_n, \bar{t}_n)$ on $\mathfrak{D}$ restricted to the function $\mathcal{F}_{\text{Homeo}_{C}(S^1)}(\mathrm{t}_n) = \mathcal{F}_{\mathfrak{D}}\left(\mathrm{t}_n, \mathcal{Z}_n ( t_m)\right)$ on $\text{Homeo}_{C}(S^1)$. The (complex) vector fields
\begin{equation}\label{eq4_29_7}
\frac{\pa}{\pa \mathrm{t}_n} = \frac{\pa}{\pa t_n} + \sum_{m=-\infty}^{\infty}\frac{\pa \mathcal{Z}_m}{\pa t_n}\frac{\pa}{\pa \bar{t}_m}, \hspace{1cm}n\in \Z,
\end{equation}are then well-defined vector fields on the subspace $\text{Homeo}_{C}(S^1)$. Now using the same reasoning as in Section \ref{s1}, one can prove that \eqref{eq4_29_1} still holds. It follows that with $\mL=g$ and $\tilde{\mL}=f$, their evolutions with respect to $\mathrm{t}_n$ satisfy the dispersionless Toda hierarchy \eqref{Lax}. One should take note that the $\mathrm{t}_n$-flow on $\text{Homeo}_{C}(S^1)$ is different from the $t_n$-flow on $\mathfrak{D}$. We would also like to remark that now the variables $\mathrm{t}_n$ cannot be treated as local coordinates on $\text{Homeo}_{C}(S^1)$ since they are complex variables and their complex conjugates satisfy some nontrivial relations $\bar{t}_n =\mathcal{Z}_n (t_m)$ on $\text{Homeo}_{C}(S^1)$.

\section{Riemann mappings and dispersionless Toda hierarchy}

Since the functions $f^{-1}$ and $g^{-1}$ are respectively the Riemann mappings of the domains $\Omega_1^+$ and $\Omega_2^-$, it is natural to ask whether we can describe the evolutions of the Riemann mappings $(g^{-1}, f^{-1})$ by dispersionless Toda flows. Here we are not going to explore all the possibilities. We restrict our consideration to the solutions governed by the same Riemann-Hilbert data \eqref{eq4_29_3}. Formally, one can just replace all the $f$ and $g$ in  the definitions and proofs above by $f^{-1}$ and $g^{-1}$ and get the desired results. However, analytically this is not feasible. Tracing from the beginning the definitions of $t_n$ and $v_n$, we immediately bumped into the problem   that $f^{-1}\circ g$ and $g^{-1}\circ f$ are not well defined for general $f$ and $g$. To make $f^{-1}\circ g$ and $g^{-1}\circ f$   well defined, we have to restrict our consideration to the space $\text{Homeo}_{C}(S^1)$ where $f(S^1)=g(S^1)$. $g^{-1}\circ f$ and $f^{-1}\circ g$ are then $C^1$ homeomorphisms of the unit circle. In this case, the functions $\mathsf{t}_n$ and $\mathsf{v}_n$ defined as in \eqref{eq4_12_6} can be considered as Fourier coefficients. More precisely, given   a  $C^1$ homeomorphism $\gamma$ of the unit circle,   the functions $\mathsf{t}_n,
\mathsf{v}_n$, $n< 0$ , and $\mathsf{t}_0$  on $\text{Homeo}_C(S^1)$ are the  coefficients of the absolutely convergent Fourier series
expansion of $\gamma=g^{-1}\circ f$ on $S^1$:
\begin{align}\label{eq8_4_1}
\gamma(w) = -\sum_{n=1}^{\infty} n\mathsf{t}_{-n} w^{-n+1} + \mathsf{t}_0 w
-\sum_{n=1}^{\infty} \mathsf{v}_{-n} w^{n+1} , \hspace{1cm} w=e^{i\theta}.
\end{align}
   For $n >0$, the functions $\mathsf{t}_n , \mathsf{v}_n$
are   the coefficients of the Fourier series expansion of $1/\gamma^{-1}=
(1/f^{-1})\circ g$ on $S^1$:
\begin{align}\label{eq8_4_2}
\frac{1}{\gamma^{-1}(w)} =\sum_{n=1}^{\infty}n\mathsf{t}_nw^{n-1} + c_0
w^{-1} + \sum_{n=1}^{\infty} \mathsf{v}_n w^{-n-1},\hspace{1cm}
w=e^{i\theta}.
\end{align}
For the coefficient $c_0$, it is easy to check that it coincides with $\mathsf{t}_0$:
\begin{align*}
c_0 =\frac{1}{2\pi i} \oint_{S^1} \frac{1}{\gamma^{-1}(w)} dw=
\frac{1}{2\pi i} \oint_{S^1} \frac{1}{w}d\gamma(w)=\frac{1}{2\pi
i}\oint_{S^1}\frac{\gamma(w)}{w^2}dw =\mathsf{t}_0.
\end{align*}
Finally, similar to \eqref{eq4_12_7},  the function $\mathsf{\mathsf{v}}_0$ is defined  as
\begin{align*}
\mathsf{v}_0 =\frac{1}{2\pi i} \oint_{S^1}\left(\left( \log
\frac{f(w)}{w}\right) \frac{\gamma(w)}{w^2}
-\left(\log\frac{g(w)}{w}\right) \frac{1}{\gamma^{-1}(w)}\right)
dw-\frac{1}{2\pi i} \oint_{\mathcal{C}} \frac{g^{-1}(z)}{f^{-1}(z)}
\frac{dz}{z}.
\end{align*}
Heuristically, it is equal to
\begin{align*}
\mathsf{v}_0 =-\frac{1}{2\pi i} \oint_{S^1}\left((\log w)
\frac{\gamma(w)}{w^2} - \frac{\log w}{\gamma^{-1}(w)}\right) dw.
\end{align*}

One should take note that as in Section \ref{s1}, the condition $f(S^1)=g(S^1)$ implies some nontrivial relations between the variables $\mathsf{t}_n, \bar{\mathsf{t}}_n, n\in \Z$. As in Section \ref{s1}, assume that locally, we can regard $\text{Homeo}_{C}(S^1)$ as a submanifold of a complex manifold defined by the zeros of the functions $\bar{t}_n - \mathfrak{Z}_n (t_m)$, then we can define the vector fields $\frac{\pa}{\pa \mathsf{t}_n}$ by   \eqref{eq4_29_7}.
Alternatively, one can regard $\frac{\pa}{\pa \mathsf{t}_n}$ as complexified vector fields on $\text{Homeo}_C(S^1)$ as shown by the following proposition.
\begin{proposition}\label{p4_29_1}
There are complexified vector fields $\pa_n$ on $\text{Homeo}_C(S^1)$ such that $\pa_n \mathsf{t}_m=\delta_{n,m}$.
\end{proposition}
\begin{proof}

At every point $\gamma\in \text{Homeo}_C(S^1)$, a local coordinate chart is given by $(u_0, \text{Re}\, u_n , \text{Im}\, u_n)\mapsto e^{i(\theta+u(\theta))}\circ \gamma$, where $u(\theta)=\sum_{n\in \Z} u_n e^{i n\theta}$ and $u_{-n}=\bar{u}_n$. Equivalently, we can also use $(u_n)_{n\in \Z}$ as local coordinates and a complexified vector field $\pa$ on $\text{Homeo}_C(S^1)$  can be written as
\begin{equation*}
\pa=\sum_{n=-\infty}^{\infty} c_n\frac{\pa}{\pa u_n},
\end{equation*}where $c_n\in \C$. Its action on $\gamma$ is
\begin{equation*}
\pa \gamma (w) =i\left( \sum_{n=-\infty}^{\infty}c_n w^{n+1}\right)\circ \gamma(w).
\end{equation*}Let $b_{m,n}$ be the generalized Grunsky coefficients of $(f,g)$ and $P_n(z), Q_n(z)$ the associated generalized Faber polynomials. Consider the complexified vector fields $\pa_n$ whose action on  $\gamma$ is given by
\begin{equation}\label{eq12_29_2}\begin{split}
\pa_n \gamma(w) =& \sum_{m=1}^{\infty} mn b_{-m, n} w^{m+1}= w^2 P_n'(f(w))f'(w),\\
\pa_{-n}\gamma(w)=& -n w^{-n+1}+ \sum_{m=1}^{\infty} mn b_{-m, -n} w^{m+1} = w^2Q_n'(f(w))f'(w),
\end{split}\end{equation}for $n\geq 1$ and
\begin{equation}\label{eq12_29_3}
\pa_0\gamma(w) = w-\sum_{m=1}^{\infty} m b_{m,0} w^{m+1} =w^2\frac{f'(w)}{f(w)}.
\end{equation}
 As in the proof of Proposition \ref{p4_12_1}, one can show that $\pa_n, n\in\Z$, give rise to independent variations of $\gamma$.

From the definitions \eqref{eq8_4_1} and \eqref{eq8_4_2}, we find that  for $n\in \Z$,
\begin{equation}\label{eq12_29_7}
\pa_{ n} \gamma(w) = -\sum_{m=1}^{\infty} m[\pa_{ n} \mathsf{t}_{-m}]w^{-m+1} +[\pa_{n} \mathsf{t}_0] w -\sum_{m=1}^{\infty}[\pa_{n} \mathsf{v}_{-m}] w^{m+1},
\end{equation}and
\begin{equation}\label{eq12_29_8}
\frac{(\gamma^{-1})'(w) (\pa_n \gamma)\circ \gamma^{-1}(w)}{\gamma^{-1}(w)^2} = \sum_{m=1}^{\infty} m[\pa_{ n} \mathsf{t}_{m}]w^{m-1} +[\pa_{n} \mathsf{t}_0 ] w^{-1} +\sum_{m=1}^{\infty}[\pa_{n} \mathsf{v}_{m}] w^{-m-1}.
\end{equation}Using the definition of $\pa_n\gamma(w)$ given by \eqref{eq12_29_2} and \eqref{eq12_29_3}, we  find from \eqref{eq12_29_7} that
\begin{equation}\label{eq12_29_4}
\sum_{m=1}^{\infty} mn b_{-m, n} w^{m+1}=-\sum_{m=1}^{\infty} m[\pa_{ n} \mathsf{t}_{-m}]w^{-m+1} +[\pa_{n} \mathsf{t}_0] w -\sum_{m=1}^{\infty}[\pa_{n} \mathsf{v}_{-m}] w^{m+1},
\end{equation}
\begin{equation}\label{eq12_29_5}\begin{split}
&-n w^{-n+1}+ \sum_{m=1}^{\infty} mn b_{-m, -n} w^{m+1}\\=&-\sum_{m=1}^{\infty} m[\pa_{ -n} \mathsf{t}_{-m}]w^{-m+1} +[\pa_{-n} \mathsf{t}_0] w -\sum_{m=1}^{\infty}[\pa_{-n} \mathsf{v}_{-m}] w^{m+1},\hspace{2cm}\end{split}
\end{equation}for $n\geq 1$, and
\begin{equation}\label{eq12_29_6}
 w-\sum_{m=1}^{\infty} m b_{m,0} w^{m+1}= -\sum_{m=1}^{\infty} m[\pa_{ 0} \mathsf{t}_{-m}]w^{-m+1} +[\pa_{0} \mathsf{t}_0] w -\sum_{m=1}^{\infty}[\pa_{0} \mathsf{v}_{-m}] w^{m+1}.
\end{equation}On the other hand, using \eqref{eq12_29_8} and $f\circ \gamma^{-1}=g$, we have
\begin{equation}\label{eq12_29_9}\begin{split}
&\sum_{m=1}^{\infty} m[\pa_{ n} \mathsf{t}_{m}]w^{m-1} +[\pa_{n} \mathsf{t}_0 ] w^{-1} +\sum_{m=1}^{\infty}[\pa_{n} \mathsf{v}_{m}] w^{-m-1}\\=&\frac{(\gamma^{-1})'(w) (w^2P_n'\circ f)\circ \gamma^{-1}(w) f'\circ \gamma^{-1}(w)}{\gamma^{-1}(w)^2}=P_n'(g(w))g'(w)\\
=& nw^{n-1}- \sum_{m=1}^{\infty} nm b_{nm}w^{-m-1};\end{split}
\end{equation}
\begin{equation}\label{eq12_29_10}\begin{split}
&\sum_{m=1}^{\infty} m[\pa_{- n} \mathsf{t}_{m}]w^{m-1} +[\pa_{-n} \mathsf{t}_0 ] w^{-1} +\sum_{m=1}^{\infty}[\pa_{-n} \mathsf{v}_{m}] w^{-m-1}\\=&Q_n'(g(w))g'(w)=   -\sum_{m=1}^{\infty} nm b_{m, -n}w^{-m-1}\end{split}
\end{equation}for $n\geq 1$; and
\begin{equation}
\label{eq12_29_11}
\sum_{m=1}^{\infty} m[\pa_{0} \mathsf{t}_{m}]w^{m-1} +[\pa_{0} \mathsf{t}_0 ] w^{-1} +\sum_{m=1}^{\infty}[\pa_{0} \mathsf{v}_{m}] w^{-m-1}=\frac{g'(w)}{g(w)}=   \sum_{m=1}^{\infty} m b_{m, 0}w^{-m-1}.
\end{equation}We can then read from \eqref{eq12_29_4}, \eqref{eq12_29_5}, \eqref{eq12_29_6}, \eqref{eq12_29_9}, \eqref{eq12_29_10} and \eqref{eq12_29_11} that $\pa_n \mathsf{t}_m =\delta_{n,m}$ for all $n,m\in\Z$.
\end{proof}It follows from this proposition that we can identify the vector field $\frac{\pa}{\pa \mathsf{t}_n}$ with $\pa_n$ defined in the proof. One can also trace from the proof that
\begin{proposition}\label{Prop2}
Let $b_{m,n}$ be the generalized Grunsky coefficients of the pair
of univalent functions $(f,g)$. The variation of $\mathsf{v}_m$, $m\in \Z$, with respect to $\mathsf{t}_n$, $n\in \Z$, is given by the
following:
\begin{equation*}
\frac{\pa \mathsf{v}_m}{\pa \mathsf{t}_n} = -|mn| b_{n,m}, \hspace{0.3cm}n\neq 0,
\hspace{1cm} \text{and} \hspace{1cm} \frac{\pa \mathsf{v}_m}{\pa \mathsf{t}_0} = |m|
b_{0,m}.
\end{equation*}

\end{proposition}
For the function $\mathsf{v}_0$, we have
\begin{proposition}\label{Prop3}
The variation of $\mathsf{v}_0$ with respect to $\mathsf{t}_n$, $n\in \Z$, is given
by
\begin{align*}
\frac{\pa \mathsf{v}_0}{\pa \mathsf{t}_n} = |n|b_{n,0}=\frac{\pa \mathsf{v}_n}{\pa \mathsf{t}_0},
\hspace{0.3cm}n\neq 0, \hspace{1cm} \frac{\pa \mathsf{v}_0}{\pa \mathsf{t}_0} =
 -2b_{0,0}.
\end{align*}
\end{proposition}
\begin{proof}
We have
\begin{align*}
\frac{\pa \mathsf{v}_0}{\pa \mathsf{t}_n}=&\frac{1}{2\pi i}\oint_{S^1} \left( \log
\frac{f(w)}{w}\right)\frac{(\pa \gamma/\pa
\mathsf{t}_n)(w)}{w^2}-\left(\log\frac{g(w)}{w}\right)\left( \frac{\pa}{\pa
\mathsf{t}_n}\frac{1}{\gamma^{-1}} \right)(w) dw.
\end{align*}
Using the series expansion for each term give the desired result.
\end{proof}

 For the tau function, we define it as\begin{equation*}\begin{split}
\log \tau = \frac{\mathsf{t}_0 \mathsf{v}_0}{2} -\frac{\mathsf{t}_0^2 }{4}&+\frac{1}{8\pi i}\oint_{S^1}
\frac{1}{\gamma^{-1}(w)}\left(w\phi'(w)+
2\phi(w)\right) dw \\
&+\frac{1}{8\pi i}\oint_{S^1} \frac{\gamma(w)}{w^2}\left(w\psi'(w)
- 2\psi(w)\right) dw,\end{split}
\end{equation*}where
\begin{equation*}\begin{split}
\psi(w)&=\sum_{n=1}^{\infty} \frac{\mathsf{v}_{-n}}{n} w^n,\hspace{0.5cm}
w\in \mathbb{D},
\hspace{1.5cm}\phi(w)=\sum_{n=1}^{\infty}\frac{\mathsf{v}_n}{n} w^{-n},
\hspace{0.5cm} w\in \mathbb{D}^*.
\end{split}\end{equation*}

From Proposition \ref{Prop2} and the identities in
\eqref{iden1}, we find that the variations of $\psi(w)$ and $\phi(w)$ with respect to $\mathsf{t}_n, n\in\Z,$ are given by:
\begin{lemma}\label{lemma2}
The variations of the functions $\psi$ and $\phi$ with respect to
$\mathsf{t}_n$, $\mathsf{t}_{-n}$, $n\geq 1 $, and $\mathsf{t}_0$ are given by
\begin{align*}
\frac{\pa \psi}{\pa \mathsf{t}_n}(w)&=-P_n(f(w)) + n b_{n,0}, \hspace{1.3cm}
\frac{\pa \phi}{\pa \mathsf{t}_n}(w) =-P_n(g(w))+ w^n,\\
\frac{\pa \psi}{\pa \mathsf{t}_0}(w)&= -\log \frac{f(w)}{w}+\log
a_1,\hspace{1.5cm}\frac{\pa
\phi}{\pa \mathsf{t}_0}(w) = -\log \frac{g(w)}{w}+\log b,\\
\frac{\pa \psi}{\pa \mathsf{t}_{-n}}(w)&= -Q_n(f(w))+
w^{-n},\hspace{1.5cm}\frac{\pa \phi}{\pa
\mathsf{t}_{-n}}(w)=-Q_n(g(w))-nb_{-n,0}.
\end{align*}
\end{lemma}

From this, we can prove as in Proposition \ref{p4_27_1} that
\begin{proposition}\label{thm1}
The tau function generates the functions $\mathsf{v}_n$, namely
\begin{align*}
\frac{\pa \log \tau}{\pa \mathsf{t}_n}= \mathsf{v}_n.
\end{align*}
for all $n\in \Z$.
\end{proposition}

Combining this proposition with Proposition \ref{Prop2} and Proposition \ref{Prop3}, we have
\begin{align}\label{second}
\frac{\pa^2\log\tau}{\pa \mathsf{t}_m\pa \mathsf{t}_n}
=\begin{cases}-|mn|b_{m,n},\hspace{1cm}&\text{if}\;\;m\neq 0, n\neq 0\\
|m| b_{m,0}, &\text{if}\;\;m\neq 0,n= 0\\
-2b_{0,0},&\text{if}\;\;m=n=0.
\end{cases}
\end{align}Therefore, we conclude
by Proposition \ref{Hirota} that
\begin{theorem}\label{th1}
The evolutions of the Riemann mappings
$( g^{-1}, f^{-1})$ with respect to $\mathsf{t}_n, n\in\Z,$ satisfy the dispersionless Toda
hierarchy \eqref{Lax}.\end{theorem}

 \vspace{0.2cm} \noindent \textbf{Acknowledgement}\;
The author would like to thank A. Zabrodin and L. Takhtajan for the helpful comments. We would also like to thank the anonymous referee for the illuminating suggestions which have greatly improved the presentation of this article. This project is funded by  Ministry of Science, Technology
and Innovation of Malaysia under
eScienceFund 06-02-01-SF0021.

\appendix
\section{The subgroup of linear fractional transformations}\label{a1}

In this section, we consider the subspace of $\mathfrak{D}$ containing those $(f,g)$  where $f$ and $g$ are linear fractional transformations. The
   conditions $f(0)=0$, $g(\infty)$, $f'(0)g'(\infty)=1$, $\infty\notin f(\Del)$ and $0\notin g(\Del^*)$ imply that $f$ and $g$ have the following forms:
\begin{align*}
f(w) = \frac{1}{b}\frac{w}{1+aw}, \hspace{0.5cm} |a|<1,\hspace{1cm} g(w) = b
w+c,\hspace{0.5cm} b\neq 0, \;\left|\frac{c}{b}\right|<1.
\end{align*}Notice that here we have three complex parameters $a, b$ and $c$. It is straightforward to compute from the definition \eqref{eq4_12_6} of $t_n$ and $v_n$ that
\begin{equation}\label{eq4_29_8}
t_{-1}=-c, \hspace{0.5cm} t_0=b^2, \hspace{0.5cm} t_1= ab, \hspace{0.5cm} t_n =0 \;\;\text{for all}\; |n|\geq 2;
\end{equation}\begin{equation*}\begin{split}
\text{for}\; n\geq 1, \;\;\; v_n=b^2c^n, \hspace{0.5cm} v_{-n}=-b^{n+2}a^n, \hspace{0.5cm}\text{and}\hspace{0.5cm} v_0= b^2\log b^2-b^2+abc.
\end{split}
\end{equation*}Therefore, we see that the subspace of linear fractional transformations is characterized by $t_n =0$ for all $|n|\geq 2$. As a function of $t_{-1}, t_0$ and $t_1$, we have
\begin{equation*}
v_n =(-1)^n t_0t_{-1}^n, \hspace{0.5cm} v_{-n}= -t_0t_1^n, \hspace{0.5cm} v_0=t_0\log t_0 -t_0-t_{-1}t_1.
\end{equation*}A straightforward computation shows that the $\tau$ function \eqref{eq5_1_1} is given by
\begin{equation*}
\tau  = \left|\mathfrak{T}\right|^2=\left|\exp\left(\frac{t_0^2}{4}\log t_0^2 -\frac{3}{4} t_0^2 -t_{-1}t_{0}t_1 \right)\right|^2.
\end{equation*}From this, it is easy to verify that
\begin{equation}\label{eq4_29_9}
\frac{\pa\log\tau}{\pa t_{-1}}=v_{-1}, \hspace{0.5cm}\frac{\pa\log\tau}{\pa t_{0}}=v_{0},\hspace{0.5cm}\frac{\pa\log\tau}{\pa t_{1}}=v_{1}.
\end{equation}

Now we consider the restriction of $(f,g)$ considered above to the subspace $\Sigma$ and $\text{Homeo}_{C}(S^1)$. Restricted to $\Sigma$,
\begin{equation*}\frac{1}{b}\frac{w}{1+aw}=f(w)=\frac{1}{\overline{g(1/\bar{w})}}=\frac{1}{\bar{b}}\frac{w}{1+\frac{\bar{c}}{\bar{b}}w}.\end{equation*}
Therefore, $$b=\bar{b}\hspace{0.5cm}\text{and}\hspace{0.5cm} \frac{\bar{c}}{\bar{b}}=a.$$\eqref{eq4_29_8} then implies that $t_0$ is real and $\bar{t}_1=-t_{-1}$. Therefore
\begin{equation*}
\bar{v}_n =t_0t_1^n=-v_{-n}\;\;\;\;\text{for}\;\; n\neq 1,\hspace{0.5cm}\text{and}\hspace{0.5cm} v_0=t_0\log t_0-t_0+|t_1|^2,
\end{equation*}and the tau function $\mathfrak{T}$ is
\begin{equation*}
\mathfrak{T}=\exp\left(\frac{t_0^2}{4}\log t_0^2 -\frac{3}{4} t_0^2 +t_0|t_1|^2 \right).
\end{equation*}Again, one can show that \eqref{eq4_29_9} holds.

For the restriction to $\text{Homeo}_{C}(S^1)$, the condition $f(S^1)=g(S^1)$ is equivalent to $\gamma=g^{-1}\circ f$ is a linear fractional transformation mapping $S^1$ to itself. Equivalently, $\gamma\in\PSL(2,\R)$. This implies that
\begin{equation*}
b = \frac{e^{\frac{i\alpha}{2}}}{\sqrt{ (1-|a|^2)}},  \;\;c= -\frac{\bar{a}e^{-\frac{i\alpha}{2}}}{\sqrt{ (1-|a|^2)}},
\hspace{0.5cm}\text{and}\hspace{0.5cm} \gamma(w)=
e^{-i\alpha}\frac{w+\bar{a}}{1+aw}.
\end{equation*}Substituting into \eqref{eq4_29_8} gives
\begin{equation*}\begin{split}
 t_{-1}=\frac{\bar{a}e^{-\frac{i\alpha}{2}}}{\sqrt{1-|a|^2}}, \hspace{0.5cm} t_0=\frac{e^{i\alpha}}{1-|a|^2}, \hspace{0.5cm} t_1=\frac{ae^{\frac{i\alpha}{2}}}{\sqrt{1-|a|^2}}.\end{split}
\end{equation*}Therefore,
\begin{equation*}
\bar{t}_{-1}=t_1, \hspace{0.5cm}\bar{t_0}= \frac{(1+t_1t_{-1})^2}{t_0}, \hspace{0.5cm}\bar{t}_1=t_{-1}.
\end{equation*}As a function of $t_{-1}, t_0, t_1$, the tau function is given by
\begin{equation}\label{eq4_29_10}
\mathfrak{T}=\exp\left(\frac{t_0^2}{4}\log t_0^2 -\frac{3}{4} t_0^2 -t_{-1}t_0t_1  \right).
\end{equation}\eqref{eq4_29_9} still holds.

Next we consider the time variables $\mathsf{t}_n$ and the functions $\mathsf{v}_n$ for the evolutions of the Riemann mappings $(g^{-1}, f^{-1})$. Since \begin{align*}
\gamma(w) = \bar{a}e^{-i\alpha} +\sum_{n=0}^{\infty}(-1)^n
e^{-i\alpha}
a^{n} (1-|a|^2) w^{n+1},\\
 \frac{1}{\gamma^{-1}}(w) =- a+ \sum_{n=0}^{\infty}
 e^{-i(n+1)\alpha}(1-|a|^2)\bar{a}^nw^{-n-1},
\end{align*}
we find that the variables $\mathsf{t}_n$ and $\mathsf{v}_n$ are given by
\begin{align*}
\mathsf{t}_{1}&=- a, \hspace{0.5cm}\mathsf{t}_0= e^{-i\alpha}(1-|a|^2),\hspace{0.5cm}  \mathsf{t}_{-1}=-\bar{a}
e^{-i\alpha},\;\hspace{0.5cm} \mathsf{t}_n=0, \hspace{0.5cm}\forall |n|\geq
2,\\
\mathsf{v}_n &=
 e^{-i(n+1)\alpha}(1-|a|^2)\bar{a}^n, \hspace{0.3cm} \mathsf{v}_{-n}=(-1)^{n-1}e^{-i\alpha}
a^{n} (1-|a|^2), \hspace{0.5cm}n\geq 1.
\end{align*}On the other hand, we also have $$ \mathsf{v}_0 = \mathsf{t}_0 \log \mathsf{t}_0 -\mathsf{t}_0-\mathsf{t}_1\mathsf{t}_{-1}.$$
The local coordinates $\alpha, a, \bar{a}$  of $\text{Homeo}_C(S^1)$ can be expressed in terms of $\mathsf{t}_{-1}, \mathsf{t}_0$ and $\mathsf{t}_{1}$ by
\begin{align*}
a=-\mathsf{t}_1,\hspace{0.5cm}e^{-i\alpha}=\mathsf{t}_0 + \mathsf{t}_1\mathsf{t}_{-1},\hspace{0.5cm}\bar{a}=-\frac{\mathsf{t}_{-1}}{\mathsf{t}_0+\mathsf{t}_1 \mathsf{t}_{-1}},
\end{align*}and the functions $\mathsf{v}_n, n\in \Z$ can be written as functions of $\mathsf{t}_{-1}, \mathsf{t}_0$ and $\mathsf{t}_{1}$ by
\begin{align*}
\mathsf{v}_n&=(-1)^n \mathsf{t}_0\mathsf{t}_{-1}^n, \hspace{0.5cm} \mathsf{v}_{-n}= -\mathsf{t}_0\mathsf{t}_1^n, \hspace{0.5cm}
 \mathsf{v}_0 = \mathsf{t}_0 \log \mathsf{t}_0 -\mathsf{t}_0-\mathsf{t}_1\mathsf{t}_{-1}.
\end{align*} In terms of $\mathsf{t}_{-1}, \mathsf{t}_0$ and $\mathsf{t}_{1}$, we have\begin{align*}
\bar{\mathsf{t}}_{-1}=& \frac{\mathsf{t}_1}{\mathsf{t}_0+\mathsf{t}_1\mathsf{t}_{-1}},\hspace{0.5cm}\bar{\mathsf{t}}_0  = \frac{\mathsf{t}_0}{(\mathsf{t}_0+\mathsf{t}_1\mathsf{t}_{-1})^2},\hspace{0.5cm}\bar{\mathsf{t}}_1 =    \frac{\mathsf{t}_{-1}}{\mathsf{t}_0+\mathsf{t}_1 \mathsf{t}_{-1}} .
\end{align*}
The tau function is given by
\begin{align*}
 \tau =\exp\left( \frac{\mathsf{t}_0^2}{4} \log \mathsf{t}_0^2 -\frac{3}{4}
\mathsf{t}_0^2-\mathsf{t}_{-1}\mathsf{t}_0\mathsf{t}_{1}\right).
\end{align*}Its dependence on $\mathsf{t}_{-1}, \mathsf{t}_0, \mathsf{t}_1$ is the same as \eqref{eq4_29_10}.


\begin{thebibliography}{10}
\bibitem{12}
L.~M. Alonso, \emph{Genus-zero Whitham hierarchies in conformal-map dynamics}, Phys. Lett. B. \textbf{641} (2006), 466--473.

\bibitem{6}
L.~M. Alonso and E. Medina, \emph{Solutions of the dispersionless Toda hierarchy constrained by string equations}, J. Phys. A: Math. Gen. \textbf{37} (2004), 12005--12017.

\bibitem{7}
L.~M. Alonso and E. Medina, \emph{Exact solutions of integrable 2D, contour dynamics}, Phys. Lett. B \textbf{610} (2005), 277--282.

\bibitem{10}
L. M. Alonso, E. Medina and M. Manas, \emph{String equations in Whitham hierarchies: tau-functions and Virasoro constraints}, J. Math. Phys. \textbf{47} (2006), 083512.

\bibitem{11}
M. Bauer and D. Bernard, \emph{2D growth processes: SLE and Loewner chains}, Phys. Rep. \textbf{432} (2006), 115--221.

\bibitem{1}
A. Boyarsky, A. Marshakov, O. Ruchayskiy, P. Wiegmann and A. Zabrodin, \emph{Associativity equations in dispersionless integrable hierarchies}, Phys. Lett. B \textbf{515} (2001), 483--492.

\bibitem{8}
D. Crowdy, \emph{The Benney hierarchy and the Dirichlet boundary problem in two dimensions}, Phys. Lett. A \textbf{343} (2005), 319--329.

\bibitem{Duren}
P.~L. Duren, \emph{Univalent functions}, Grundlehren der Mathematischen
  Wissenschaften [Fundamental Principles of Mathematical Sciences], vol. 259,
  Springer-Verlag, New York, 1983.

\bibitem{Gar}
F.~P.~ Gardiner and N. Lakic, \emph{Quasiconformal {T}eichm\"uller
  theory}, Mathematical Surveys and Monographs, vol.~76, American Mathematical
  Society, Providence, RI, 2000.

\bibitem{GarSu}
F.~P. Gardiner and D.~P. Sullivan, \emph{Symmetric structures on a
  closed curve}, Amer. J. Math. \textbf{114} (1992), no.~4, 683--736.

\bibitem{Ki}
A.~A. Kirillov, \emph{K\"ahler structure on the ${K}$-orbits of a group of
  diffeomorphisms of the circle}, Funktsional. Anal. i Prilozhen. \textbf{21}
  (1987), no.~2, 42--45.

\bibitem{KY}
A.~A. Kirillov and D.~V. Yuriev, \emph{K\"ahler geometry of the
  infinite-dimensional homogeneous space ${M}={\rm {d}iff}\sb +({S}\sp 1)/{\rm
  {r}ot}({S}\sp 1)$}, Funktsional. Anal. i Prilozhen. \textbf{21} (1987),
  no.~4, 35--46, 96.

\bibitem{2}
B. Konopelchenko, L.~M. Alonso and O. Ragnisco, \emph{The partial derivative-approach to the dispersionless KP hierarchy}, J. Phys. A: Math. Gen. \textbf{34} (2001), 10209--10217.

\bibitem{4}
B. Konopelchenko and L.~M. Alonso, \emph{Dispersionless scalar integrable hierarchies, Whitham hierarchy, and the quasiclassical $\bar{\partial}$--dressing method}, J. Math. Phys. \textbf{43} (2002), 3807--3823.

\bibitem{5}
B. Konopelchenko and L.~M. Alonso, \emph{Nonlinear dynamics on the plane and integrable hierarchies of infinitesimal deformations}, Stud. Appl. Math. \textbf{109} (2002), 313--336.

\bibitem{3}
I. K. Kostov, \emph{String equation for string theory on a circle}, Nucl. Phys. B \textbf{624} (2002), 146--162.

\bibitem{KKMWZ}
I.~K. Kostov, I.~M. Krichever, M.~Mineev-Weinstein, A.~Zabrodin, and P.~B.
  Wiegmann, \emph{The {$\tau$}-function for analytic curves}, Random matrix
  models and their applications, Math. Sci. Res. Inst. Publ., vol.~40,
  Cambridge Univ. Press, Cambridge, 2001, pp.~285--299.


\bibitem{8}
I. Krichever, A. Marshakov and  A. Zabrodin, \emph{Integrable structure of the dirichlet boundary problem in multiply-connected domains}, Comm. Math. Phys. \textbf{259} (2005), 1--44.


\bibitem{Lehto}
O. Lehto, \emph{Univalent functions and {T}eichm\"uller spaces}, Graduate
  Texts in Mathematics, vol. 109, Springer-Verlag, New York, 1987.

\bibitem{MWZ}
A.~Marshakov, P.~Wiegmann and A.~Zabrodin, \emph{Integrable structure of the
  {D}irichlet boundary problem in two dimensions}, Comm. Math. Phys.
  \textbf{227} (2002), no.~1, 131--153.

\bibitem{Pom}
C.~Pommerenke, \emph{Univalent functions}, Vandenhoeck \& Ruprecht,
  G\"ottingen, 1975, With a chapter on quadratic differentials by Gerd Jensen,
  Studia Mathematica/Mathematische Lehrb\"ucher, Band XXV.

\bibitem{9}
D. Prokhorov and A. Vasil'ev, \emph{Univalent functions and integrable systems}, Comm. Math. Phys. \textbf{262} (2006), 393--410.

\bibitem{TT3}
K.~Takasaki, \emph{Dispersionless {T}oda hierarchy and two-dimensional string
  theory}, Comm. Math. Phys. \textbf{170} (1995), no.~1, 101--116.



\bibitem{TT5}
K.~Takasaki and T.~Takebe, \emph{{${\rm SDiff}(2)$} {T}oda
  equation---hierarchy, tau function, and symmetries}, Lett. Math. Phys.
  \textbf{23} (1991), no.~3, 205--214.

\bibitem{TT1}
K.~Takasaki and T.~Takebe, \emph{Integrable hierarchies and dispersionless
  limit}, Rev. Math. Phys. \textbf{7} (1995), no.~5, 743--808.

\bibitem{LT}
L.~A. Takhtajan, \emph{Free bosons and tau-functions for compact {R}iemann
  surfaces and closed smooth {J}ordan curves. {C}urrent correlation functions},
  Lett. Math. Phys. \textbf{56} (2001), no.~3, 181--228, EuroConf\'erence
  Mosh\'e Flato 2000, Part III (Dijon).

\bibitem{LT2}
L.~A. Takhtajan and L.~P. Teo, \emph{Weil-Petersson metric on the universal Teichmuller space}, Mem. Amer. Math. Soc., \textbf{183} (2006), no. 861, vi+119.

\bibitem{Teo03}
L.~P. Teo, \emph{Analytic functions and integrable
  hierarchies---characterization of tau functions}, Lett. Math. Phys.
  \textbf{64} (2003), no.~1, 75--92.

\bibitem{Teo04}
L.~P. Teo, \emph{The {V}elling-{K}irillov metric on the universal {T}eichm\"uller
  curve}, J. Anal. Math. \textbf{93} (2004), 271--307.

\bibitem{UT}
K, Ueno and K. Takasaki, \emph{Toda lattice hierarchy},
Group
  representations and systems of differential equations (Tokyo, 1982), Adv.
  Stud. Pure Math., vol.~4, North-Holland, Amsterdam, 1984, pp.~1--95.
\bibitem{WZ}
P.~B. Wiegmann and A.~Zabrodin, \emph{Conformal maps and integrable
  hierarchies}, Comm. Math. Phys. \textbf{213} (2000), no.~3, 523--538.

\bibitem{Z}
A.~V. Zabrodin, \emph{The dispersionless limit of the {H}irota equations in
  some problems of complex analysis}, Teoret. Mat. Fiz. \textbf{129} (2001),
  no.~2, 239--257.

\bibitem{13}
A.~V. Zabrodin, \emph{Growth processes related to the dispersionless Lax equations}, Physica D \textbf{235} (2007), 101--108.

\end{thebibliography}
\end{document}